\newif\ifapj	
\apjtrue

\ifapj
  \documentclass[apj]{emulateapj}
\else
  \documentclass[12pt,preprint]{aastex}
\fi

\ifapj
  \usepackage{apjfonts}
\fi
\usepackage{natbib}
\usepackage[normalem]{ulem}
\usepackage{hyperref}
\shortauthors{Adams, Kochanek, Beacom, Vagins, \& Stanek}
\shorttitle{Next Galactic Supernova}

\begin{document}

\title{Observing the Next Galactic Supernova}

\author{Scott M. Adams\altaffilmark{1}, C.S. Kochanek\altaffilmark{1,2}, John F. Beacom\altaffilmark{1,2,3}, Mark R. Vagins\altaffilmark{4,5}, \& K.Z. Stanek\altaffilmark{1,2}}

\altaffiltext{1}{Dept.\ of Astronomy, The Ohio State University, 140 W.\ 18th Ave., Columbus, OH 43210}
\altaffiltext{2}{Center for Cosmology and AstroParticle Physics (CCAPP), The Ohio State University, 191 W.\ Woodruff Ave., Columbus, OH 43210}
\altaffiltext{3}{Dept.\ of Physics, The Ohio State University, 191 W.\ Woodruff Ave., Columbus, OH 43210}
\altaffiltext{4}{Kavli Institute for the Physics and Mathematics of the Universe (WPI),
Todai Institutes for Advanced Study, University of Tokyo, 5-1-5
Kashiwanoha, Kashiwa, Chiba 277-8583 Japan}
\altaffiltext{5}{Dept.\ of Physics and Astronomy, University of California, Irvine, 4129 Reines Hall, Irvine, CA 92697}

\begin{abstract}
No supernova in the Milky Way has been observed since the invention of the optical telescope, instruments for other wavelengths, neutrino detectors, or gravitational wave observatories.  It would be a tragedy to miss the opportunity to fully characterize the next one.  
To aid preparations for its observations, we model the distance, extinction, and magnitude probability distributions of a successful Galactic core-collapse supernova (ccSN), its shock breakout radiation, and its massive star progenitor.  We find, at very high probability ($\simeq 100\%$), that the next Galactic supernova will easily be detectable in the near-IR and that near-IR photometry of the progenitor star very likely ($\simeq 92\%$) already exists in the 2MASS survey.  Most ccSNe ($98\%$) will be easily observed in the optical, but a significant fraction ($43\%$) will lack observations of the progenitor due to a combination of survey sensitivity and confusion.  If neutrino detection experiments can quickly disseminate a likely position ($\sim3^{\circ}$), we show that a modestly priced IR camera system can probably detect the shock breakout radiation pulse even in daytime (64\% for the cheapest design).  Neutrino experiments should seriously consider adding such systems, both for their scientific return and as an added and internal layer of protection against false triggers.  We find that shock breakouts from failed ccSNe of red supergiants may be more observable than those of successful SNe due to their lower radiation temperatures.  We review the process by which neutrinos from a Galactic core-collapse supernova would be detected and announced.  We provide new information on the EGADS system and its potential for providing instant neutrino alerts.  We also discuss the distance, extinction, and magnitude probability distributions for the next Galactic Type Ia supernova.
Based on our modeled observability, we find a Galactic core-collapse supernova rate of $3.2^{+7.3}_{-2.6}$ per century and a Galactic Type Ia supernova rate of $1.4^{+1.4}_{-0.8}$ per century for a total Galactic supernova rate of $4.6^{+7.4}_{-2.7}$ per century is needed to account for the SNe observed over the last millennium, which implies a Galactic star formation rate of $3.6^{+8.3}_{-3.0}$ M$_{\odot}$ yr$^{-1}$.
\end{abstract}


\section{Introduction}
While we observe many extragalactic supernovae \citep[SNe; e.g.,][]{sako08,drake09,law09,leaman11}, they are very indirect probes of the SN mechanism.  This near total lack of direct constraints on the mechanism contributes to the many unsolved problems about SN, in particular, why they explode at all \citep[e.g.,][]{mezzacappa05,janka12,pejcha12,burrows13}.  Supernovae in our Galaxy and its dwarf companions, while rare, enable a broad range of new probes, in particular, neutrinos \citep[e.g.,][]{thompson03,ikeda07,marek09,abbasi11,scholberg12}, gravitational waves \citep{ott09,leonor10,yakunin10,andersson13}, nuclear $\gamma$-rays \citep{gehrels87,timmes97,hungerford03}, and shock breakout (SBO) timing \citep{matzner99,kistler12}.  Of these new probes, only neutrinos and nuclear $\gamma$-rays were demonstrated with SN 1987A \citep{hirata87,bionta87,matz88,sandie88,fryxell91,mccray93}.

Neutrinos are especially important, because they reveal the physical conditions in the core at the instant of collapse.  The detection of a burst of MeV neutrinos that will be produced from a Galactic supernova can provide the answers to three important observational questions:
\begin{itemize}

\item
\emph{IF astronomers should look for a Milky Way supernova}.  A high-statistics neutrino burst would decisively indicate that a core collapse had occurred in the Milky Way or one of its dwarf companions.  The nature of the electromagnetic transient will depend on the success of the explosion, ranging from a full supernova to something weaker to perhaps something near-impossible to detect; at present, there is no ongoing optical or IR survey that guarantees rapid detection.  Contrariwise, if no neutrinos are detected, then any electromagnetic transient is not a nearby core-collapse; it might instead be a supernova impostor or a Type Ia supernova.

\item
\emph{WHEN astronomers should look}.  Neutrino detections will reveal the time of core collapse to within seconds.  In principle, an alert could be distributed that rapidly.  This would provide an early warning that could enable the detection of the SBO signal, the early supernova light curve, and any surprises about the first electromagnetic signals following core collapse.  The precise timing will also help the detection of possible gravitational-wave signals, which may be detectable for collapses with adequate deviations from spherical symmetry.  The time-integrated neutrino signal is relatively well known and should have relatively modest variations from one event to another, including the case of failed supernovae with black hole formation.

\item
\emph{WHERE astronomers should look}.  For a Milky Way core-collapse, the Super--Kamiokande detector will be able to exploit the directionality in neutrino-electron scattering to restrict the source direction to within a few degrees.  This will greatly improve the chances of successful electromagnetic searches on short timescales.  In principle, this information could be distributed in less than a minute.  This directionality will only be valuable if there is a means to quickly exploit it with wide-field instruments to first narrow the search region and then quickly follow up with more powerful instruments.
\end{itemize}

\begin{figure}
  \includegraphics[width=9.2cm, angle=0]{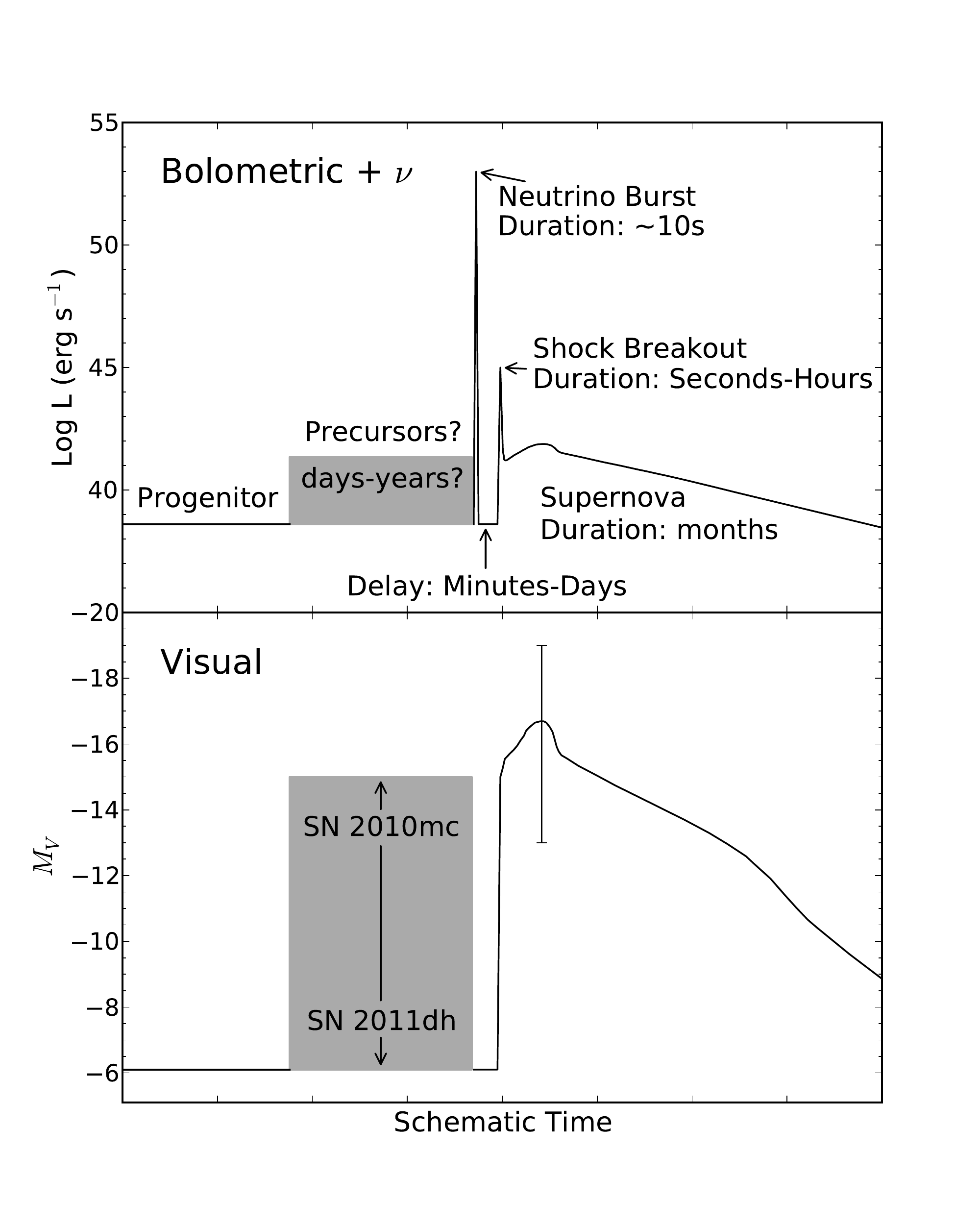}
  \caption{Schematic time sequence for the stages of a ccSN.  The scaling of the time axis varies to display vastly different timescales.  The top panel shows the combined bolometric electromagnetic and neutrino luminosities, while the bottom panel displays the typical V-band magnitudes.  The progenitor phase refers to the pre-core collapse star.  With the ignition of carbon and later stages of nuclear burning the progenitor may experience episodes of high variability in the millennia, years, or days before the core-collapse, where the maximum and minimum luminosities and magnitudes for these precursor events are from SN 2010mc \citep{ofek13} and SN2011dh \citep{szczygiel12}.  The core-collapse releases $\sim$ $10^{4}$ times more energy in neutrinos in $\sim$ 10 seconds than is released in the electromagnetic signal of the supernova over its entire duration.  The progenitor luminosity and post-shock breakout light curve are from SN1987A and its likely progenitor, SK -69d 202 \citep{arnett89,suntzeff90}, and the error bar on the peak of the $M_{V}$ light curve represents the full range of peak magnitudes observed by \cite{li11a}.  \label{fig:schematic}}
\end{figure}

Optical/near-IR observations will remain a crucial component of studies of Galactic SNe.  This includes traditional uses such as characterizing the external explosion \citep[energy, mass, composition, velocity; e.g.,][]{hamuy03} and the properties of the progenitor \citep[e.g.,][]{smartt09r}, but also new probes (see Fig. \ref{fig:schematic}), such as progenitor variability \citep[e.g.,][]{szczygiel12}, precursor eruptions \citep[e.g.,][]{pastorello07,ofek13,mauerhan13}, and constraining the existence of failed SNe \citep{kochanek08}.  Now that large neutrino detection experiments are running, the next Galactic ccSNe will also provide an unprecedented opportunity to measure the delay time between neutrino detection and shock breakout, which would probe the density structure of the progenitor \citep{kistler12}.  All these applications depend critically on the optical/near-IR observability of the SNe and its progenitor given our position near the midplane of a dust-filled disk galaxy.

Aspects of this problem have been discussed previously.  \cite{bergh75} presents predictions of the V-band observability of the next Galactic SNe assuming the Galaxy was a uniform disk with uniform absorption and a uniform incidence of SNe and further discusses the prospects of distance determination.  \cite{tammann94} use a similar exercise to infer the Galactic SN rate; their model consists of thin disk, thick disk, and halo components as well as a dust distribution, but no further details are given.

Given the improvement in models of the Galactic dust distribution, it is worth revisiting the estimates of \cite{bergh75} and \cite{tammann94}.  We model the SNe distribution with a double-exponential disk model using modern estimates of the scale lengths and heights for each population, and model the extinction with a similar double-exponential distribution normalized to the line of sight extinction of modern dust maps.  We present results for both V-band and near-IR observability.  We also fold in the observed luminosity functions of SNe and estimate the probability of identifying the SNe progenitor in archival data.

We separately consider SNe Ia and ccSNe, since they should have differing spatial distributions, and use our modeled SN observability to infer a Galactic supernova rate.  We also predict the observability of the shock breakout and failed supernovae.  Finally we review the neutrino detection process and discuss how near real-time neutrino alerts could be provided.  In \S2 we define our models.  We discuss the electromagnetic observability results in \S3, neutrino detection in \S4, and present our conclusions in \S5.  Two appendices outline observational systems to detect Galactic SBO emission even in daytime and for observing extragalactic SBO events within the Local Volume.  Throughout the paper we use the Vega magnitude system.

\section{Models}
\label{sec:models}
The basis of our model is a Monte Carlo simulation of the positions of Galactic SNe and their corresponding dust extinctions.  We model the progenitor and dust distributions using the double-exponential spatial distribution
\begin{equation}
\rho = A e^{-R/R_{d}}e^{-|z|/H}
\label{eq:exp}
\end{equation}
where $R$ is the Galactocentric radius and $z$ is the height above the Galactic mid-plane.  We must define $A$, $R_{d}$, and $H$ for the dust distribution, the core-collapse supernova (ccSN) progenitors, and the Type Ia (SN Ia) distribution.  We outline our approach for each of these cases in the following subsections.  For these models we use several of the same input parameters as TRILEGAL (TRIdimensional modeL of thE GALaxy), a population synthesis code for simulating stellar populations along any direction through the Galaxy \citep{girardi05}.  The Sun is placed $H_{\odot} = 24$ pc above the mid-plane of the disk at a Galactocentric radius of $R_{\odot} = 8.7$ kpc.  The Galactic thin and thick disk components are truncated at $R_{out} = 15$ kpc.

\subsection{Dust}
\label{sec:dust}
We assume that dust largely traces star formation, and thus use the scale length of the thin disk for the scale length of the dust distribution.  We adopt a scale length of $R_{d} = 2.9$ kpc from the TRILEGAL model.  TRILEGAL uses a scale height of $H = 110$ pc for the dust and $H = 95$ pc for the thin disk.  While we calculated results for both values, the differences were so small that we only discuss the result for $H = 110$ pc.  These choices for the spatial distribution are less critical than the estimated total extinction along any line of sight.  We separately consider four possible normalizations for the total line of sight extinction.  The simplest method we use (hereafter referred to as SIMPLE) distributes the dust following Eqn. \ref{eq:exp} and normalizes the distribution to have $A_{V} = 30$ to the Galactic center.  In the remaining models, we distribute the dust along any line of sight following Eqn. \ref{eq:exp}, but normalize each line of sight using an empirical model for the total extinction in that direction.  In our second model (SFD98), we normalize the extinction along each line of sight by the total extinction from \cite{schlegel98}.  However, \cite{schlegel98} is believed to overestimate $E(B-V)$ in regions of high extinction \citep{stanek98,arce99,chen99}.  To account for this, we consider a modified SFD98 model (modSFD98), where we correct the high extinction values following \cite{bonifacio00}, such that
\ifapj
  $E(B-V)' = E(B-V)$ for $E(B-V) \leq 0.1$ and $E(B-V)' = 0.1 + 0.65(E(B-V)-0.1)$ for $E(B-V) > 0.1$,
\else
  \begin{equation}
  E(B-V)' = \left\lbrace \begin{array}{lcl}
  E(B-V) & \mbox{for} & E(B-V) \leq 0.1 \\
  0.1 + 0.65(E(B-V)-0.1) & \mbox{for} & E(B-V) > 0.1
  \end{array}\right.,
  \label{eq:extinction_fiddle}
  \end{equation}
\fi
which significantly reduces the total extinction in the Galactic plane.  Since the SFD98 dust maps may be completely problematic in the areas of high extinction found near the Galactic midplane \citep[e.g.,][]{majewski11}, we also consider a model employing the Rayleigh-Jeans Color Excess (RJCE) extinction map of the Galactic midplane presented by \cite{nidever12} where possible, falling back to the modified SFD98
\ifapj
\else
  (Eqn. \ref{eq:extinction_fiddle})
\fi
only in the 17\% (42\%) of cases where our ccSNe (SNe Ia) lay outside of the RJCE extinction map footprint.  We note that the RJCE extinction map is derived from red giant branch stars which lie within the Galaxy, and so only estimates the total extinction out to 18-20 kpc from the observer.  This should still represent the total extinction for most of the simulated SN positions.  For comparison we also present results that assume no extinction (No Dust).
We adopt $A_{V} = R_{V} E(B-V)$ and $A_{K} = 0.114 R_{V} E(B-V)$ following \cite{cardelli89}, with $R_{V}=3.1$.  

The extinction law, the value of $R_{V}$ at its simplest, is not uniform in the Galaxy.  The value of $R_{V}=3.1$ we adopt is an average value \citep[e.g.,][]{cardelli89}.  Dense molecular clouds can have far larger values of $R_{V}$ \citep[e.g.,][]{jenniskens93,olofsson10}, but molecular clouds cover a small fraction of the sky.  Dust in lower extinction directions towards the bulge show evidence for an $R_{V}$ significantly below $R_{V} =3.1$ \citep{nataf13}, but if extinction is low, the particular value of $R_{V}$ is unimportant.  Readers should be aware of these issues, but they will not dominate our results in any typical direction where extinction is dominated by integrating the normal ISM over long sight lines.

\subsection{ccSNe}
\label{sec:ccSNe}
\begin{figure}
  \includegraphics[width=9.2cm, angle=0]{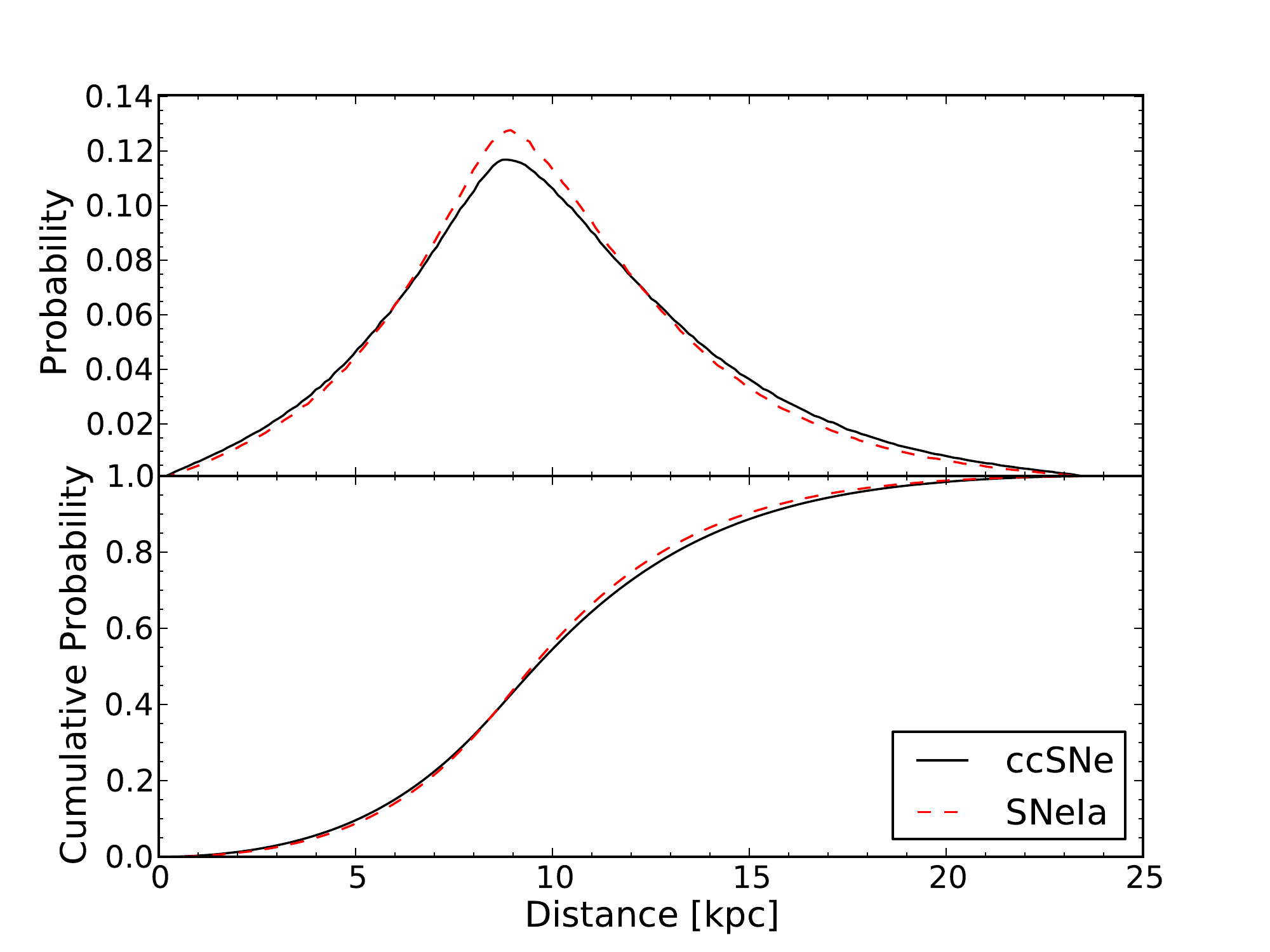}
  \caption{Differential (top) and cumulative (bottom) distance distributions of Galactic SNe from the Sun.  Reasonable changes in the distance distributions have little effect on the visibility, so we only present the fiducial case.  In particular, 3-D structure, such as spiral arms, would produce features in this figure but would have little consequence for the magnitude distribution of SNe, as discussed in \S\ref{sec:ccSNe}.\label{fig:dcomp}}
\end{figure}

\begin{figure}
  \includegraphics[width=9.2cm, angle=0]{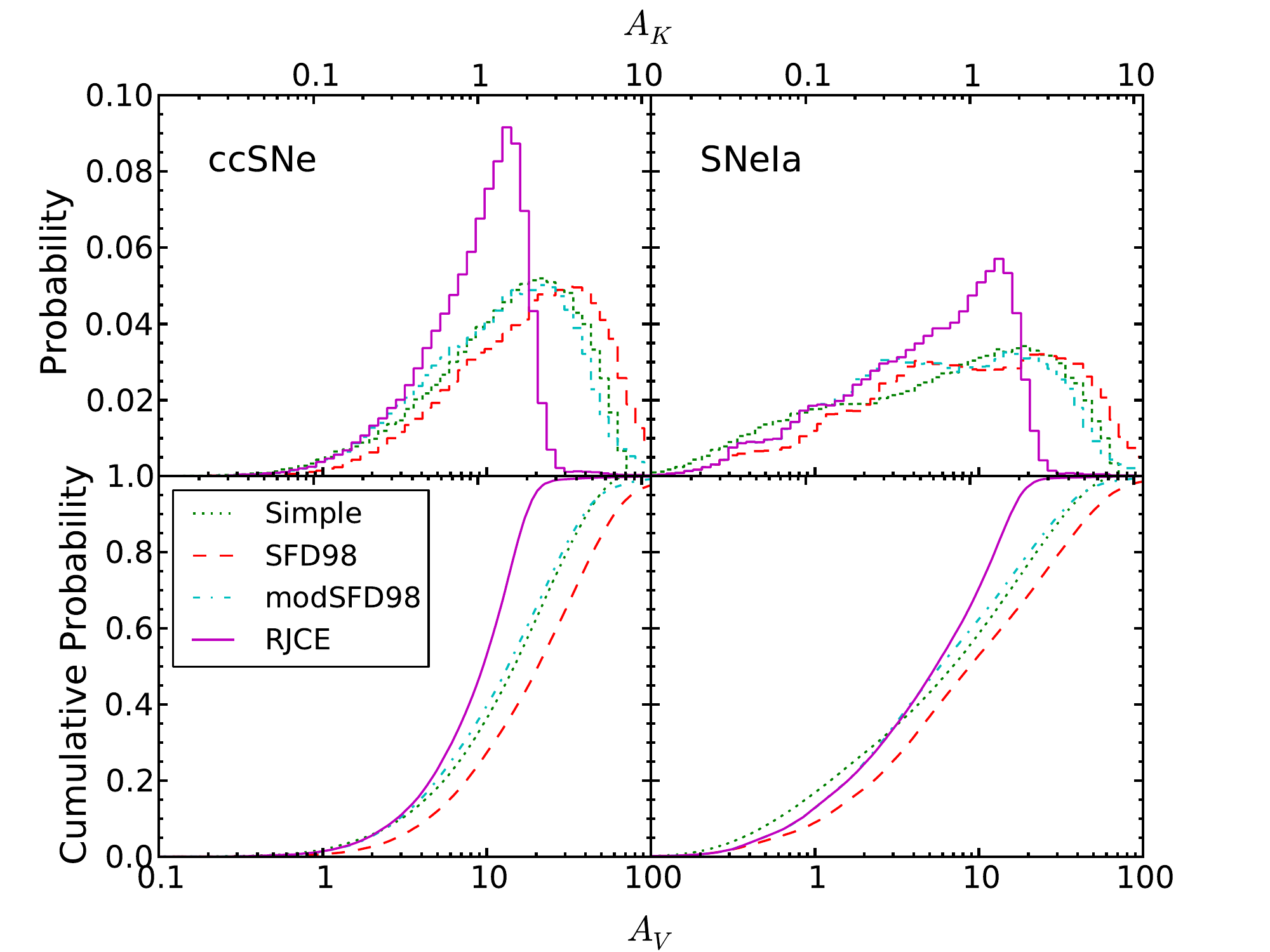}
  \caption{Differential (top) and integral (bottom) extinction distributions for ccSNe (left) and SNe Ia (right).  The bottom axis for each panel gives the V-band extinction (in magnitudes), while the top axis gives the K-band extinction.  The models for the different dust normalizations are described in \S\ref{sec:dust}.  The model dependence of the extinction distribution, rather than the distance distribution, is the primary source of uncertainty in the visibility of SNe and their progenitors.\label{fig:Acomp}}
\end{figure}

We assume that ccSNe trace the thin disk and use the thin disk parameters from TRILEGAL ($R_{d} = 2.9$ kpc and $H = 95$ pc) described in \S\ref{sec:dust}.  The distance probability distribution of ccSNe for these parameters is shown in Fig. \ref{fig:dcomp} and the extinction probability distributions of ccSNe for the different dust models are displayed in the left panels of Fig. \ref{fig:Acomp}.  We only present results for this single set of thin disk parameters because, to the extent that dust traces massive star formation, the exact choice of disk parameters is relatively unimportant.  First, if the dust distribution traces the distribution of massive star formation, then the differential distribution of ccSNe along a line of sight, $dN/dl$, is proportional to the differential of the optical depth along the line of sight, $d\tau/dl$.  Thus, if dust traces star formation, the differential distribution of the progenitors in optical depth, $dN/d\tau$, is independent of the line of sight spatial distribution chosen for the ccSNe and the dust.  Second, any effects from changing the spatial distribution are negligible compared to the differences between the extinction models.
For example, consider the effects of adding a spiral arm at 1 kpc with a 1 kpc inter-arm distance.  Our model will spread the star formation of that arm uniformly over $\sim1/2$ the inter-arm separation rather than putting it in the arm.  This means we spread the distance modulus by $\sim 0.6$ mag at most, which is negligible compared to the effects of the dust distribution.  For more distant arms, the problem rapidly becomes even smaller, so the detailed 3-D structure of the disk is unimportant to our results.

Given the distances and extinctions to each supernova position, we calculate the apparent magnitude probability distribution of ccSNe.  We first consider a case using a fixed magnitude of $M_{V,max} = -16$ and $V-K = 1.0$, where the color is a ``typical" value from \cite{krisciunas09}.  This simple case allows the reader to easily rescale the observability for arbitrary luminosity and color.  We also present the magnitude distribution obtained by folding in the ccSNe luminosity function found by \cite{li11a} and use this case for quantitative estimates of the observability of ccSNe.  While folding in the luminosity function broadens the resulting magnitude distribution only slightly, this effect is easy to include.

We find the apparent magnitude distribution for the ccSNe progenitor population by assuming that the number distribution of the population is given by a Salpeter IMF ($dN/dM \varpropto M^{-2.35}$) with a minimum mass of $8 \mathrm{M}_{\odot}$ and a maximum mass of $100 \mathrm{M}_{\odot}$.  To find the progenitor luminosity for a given mass, we rely on an interpolation of the Padova isochrones \citep{marigo08}, taking the progenitor luminosity to be the luminosity of the most massive star left on the isochrone.  Other models would yield moderately different results due to differing treatments of mass loss and the transition between being red or blue supergiants and Wolf-Rayet stars at the time of explosion \citep[e.g.,][]{groh13}.

\subsection{SNe Ia}
\cite{mannucci06} and \cite{brandt10} find that SNe Ia progenitors can be described by a bimodal progenitor delay time distribution, with approximately half the SNe Ia occurring at stellar ages of order 100 Myr and the remaining half occurring on Gyr timescales.  Therefore, we draw our SNe Ia progenitors equally from the thin disk population used for the dust and ccSNe in \S\ref{sec:dust} and \S\ref{sec:ccSNe} and from a thick disk population with $R_{d} = 2.4$ kpc and $H = 800$ pc, again following the TRILEGAL parameters.    
We note that recent work has advocated a continuous delay time distribution \citep{horiuchi10,maoz12}, but this extra complication seemed unnecessary for our present models.
As with the ccSNe, the distance and extinction probability distributions of SNe Ia are shown in Figs. \ref{fig:dcomp} and \ref{fig:Acomp}.
We give the cumulative magnitude probability density for SNe Ia of a characteristic magnitude of $M_{V} = -18.5$ and $V-K = -0.7$ \citep{folatelli10} and by folding in the SNe Ia luminosity function from \cite{li11a}, using the results of the latter for quantitative estimates of the observability of SNe Ia.  As for the ccSNe, the elaboration of including the observed luminosity function is simple but only slightly broadens the resulting magnitude distribution.

\subsection{Confusion}
The observability of SNe or their progenitors located towards the Galactic center may be reduced by confusion.  We will discuss confusion only in relation to the progenitors of ccSNe since the nature of SN Ia progenitors is debated.  We note, however, that current searches for binary companions to SN Ia in the Galaxy and LMC are primarily limited by the difficulties in inferring the position of the SN from the geometry of the remnant \citep[see, e.g.,][]{edwards12,kerzendorf12,schaefer12}.  The position of any new SN Ia would be directly measured, greatly simplifying the search for either the progenitor or a surviving companion, donor star.  To estimate the effect of confusion in the near-IR, we measured the density of sources brighter than $m_{0K}=12$ in the 2MASS catalog \citep{skrutskie06} towards each simulated progenitor position.  We then extrapolate the integrated surface density, $\Sigma_{K}$, based on a power law, $\Sigma_{K} = \Sigma_{0K}10^{\alpha_{K}(m_{prog}-m_{0K})}$, where $m_{prog}$ is the apparent magnitude of the progenitor, $\alpha_{K}$ is the power law index, and $\Sigma_{0}$ is the source density to the limiting magnitude.  We estimated $\alpha_{K} \sim 0.4$ by fitting the slope of the log of the number of sources in the 2MASS catalog for different limiting magnitudes at different coordinates.   

We followed a similar procedure for estimating the effect of confusion in the V-band.  We measured the density of sources brighter than $m_{0V}=17$ in the USNO-B1.0 catalog \citep{monet03} towards each simulated progenitor position, with rough $V$ magnitudes estimated from the photographic $R$ and $B$ magnitudes by the relation\footnote{\url{http://www.aerith.net/astro/color_conversion.html}} $V = 0.625R + 0.375B$.  We extrapolate the integrated surface density of sources in V-band by $\Sigma_{V} = \Sigma_{0V}10^{\alpha_{V}(m_{prog}-m_{0V})}$, where we have estimated that $\alpha_{V} \sim 0.45$.
While the available data are not ideal for these estimates, they should be adequate.  
For each Monte Carlo realization, the probability, $P$, of finding a source with $m<m_{target}$ within a given radius of the target is $P = 1 - e^{-\Sigma\theta^{2}}$.  

\subsection{Shock Breakout}
\label{sec:sbo}
The first electromagnetic signature from a SN is not the familiar rise to peak and then decline over weeks or months, but a short flash of radiation as the shock wave ``breaks out" from the surface of the star.  While this SBO phenomenon also occurs in SNe Ia, we focus on ccSNe because SNe Ia do not emit a (currently) detectable neutrino signal that could be used to trigger a search \citep{odrzywolek11} and the shock breakout from a white dwarf would be much fainter than that from a massive star \citep{piro10}.
The breakout pulse from ccSNe has only been observed a few times \citep[GRB 060218/SN 2006aj, XRT 080109/SN 2008D, and SNLS-04D2dc;][]{campana06,soderberg08,schawinski08} because its characteristic duration roughly corresponds to the light crossing time of the star, $R_{*}/c$.  It occurs with a delay after the neutrino or gravity wave pulse set by the time for the shock to reach the surface of the star, making it a probe of the structure of the star \citep{kistler12}.  The search for a pulse lasting seconds to hours occurring minutes to days after a neutrino or gravitational wave trigger is challenging for observers based on a rotating Earth orbiting a bright star and embedded in a dusty Galactic disk.

\begin{figure}
  \includegraphics[width=9.2cm, angle=0]{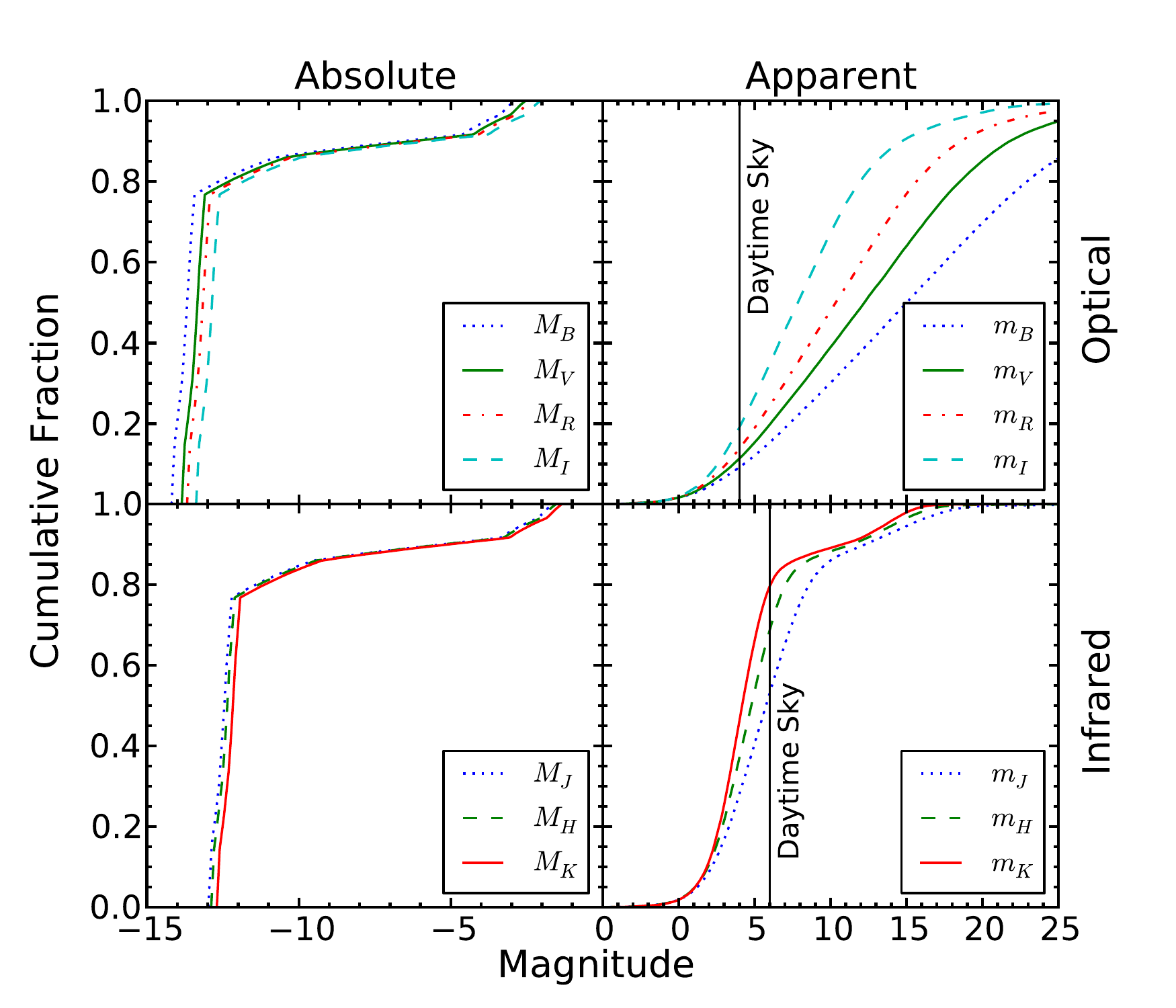}
  \caption{Cumulative absolute (left) and apparent (right) magnitude probability distributions in the optical (top) and near-IR (bottom) for the shock breakout radiation from ccSNe.  These estimates assume that the SBO radiation is thermalized.  For comparison we show the approximate surface brightness (in $\mathrm{mag}/\mathrm{arcsec}^{2}$) of the daytime sky in the visible and in the near-IR.  Note that the majority of shock breakouts from Galactic SNe appear brighter than the daytime sky in the near-IR.\label{fig:sbo_mag}}
\end{figure}

As a simple approximation for the SBO properties we adopt the $n=3$ (radiative) polytrope model of \cite{matzner99}, similar to the recent work by \cite{kistler12}.  Fixing the explosion energy at $10^{51}$ erg, simply using Thomson opacities and defining the luminosity as the characteristic energy divided by the characteristic time scale, we obtain order of magnitude estimates that
\ifapj
  \begin{eqnarray}
  T_{\mathrm{eff}} \sim 1.24 \times 10^{6} \mathrm{K} \left( \frac{M_{\star}}{M_{\odot}} \right)^{0.046} \left( \frac{M_{\mathrm{ej}}}{10 M_{\odot}} \right)^{-0.114} \nonumber \\ \times \left( \frac{R}{50 R_{\odot}} \right)^{-0.48}
  \end{eqnarray}
and
  \begin{eqnarray}
  L \sim 1.66 \times 10^{45} \mathrm{erg/s} \left( \frac{M_{\star}}{10 M_{\odot}} \right)^{0.126} \left( \frac{M_{\mathrm{ej}}}{10 M_{\odot}} \right)^{-0.816} \nonumber \\ \times \left( \frac{R}{50 R_{\odot}} \right)^{-0.22}
  \end{eqnarray},
\else
  \begin{eqnarray}
  T_{\mathrm{eff}} \sim 1.24 \times 10^{6} \mathrm{K} \left( \frac{M_{\star}}{M_{\odot}} \right)^{0.046} \left( \frac{M_{\mathrm{ej}}}{10 M_{\odot}} \right)^{-0.114} \left( \frac{R}{50 R_{\odot}} \right)^{-0.48} \nonumber
  \end{eqnarray}
and
  \begin{eqnarray}
  L \sim 1.66 \times 10^{45} \mathrm{erg/s} \left( \frac{M_{\star}}{10 M_{\odot}} \right)^{0.126} \left( \frac{M_{\mathrm{ej}}}{10 M_{\odot}} \right)^{-0.816} \left( \frac{R}{50 R_{\odot}} \right)^{-0.22},
  \end{eqnarray}
\fi
where $M_{\star}$ is the mass of the progenitor, $M_{\mathrm{ej}}=M_{\star}-1.4 M_{\odot}$ is the mass of the ejecta assuming a neutron star is formed, and $R$ is the progenitor radius.  Combined with our model for the progenitor properties (see \S \ref{sec:progenitor}), this leads to the predicted distribution of the peak absolute magnitudes of the SBO shown in Fig. \ref{fig:sbo_mag} if we assume the radiation is thermalized to a blackbody spectrum.  This is an important assumption \citep[see, e.g., the discussion in][]{nakar10}, and requires careful consideration as part of any full design study of our proposal for a Galactic SBO detection system in Appendix \ref{app:One} or extragalactic SBO detection system in Appendix \ref{app:Two}.  \cite{sapir13} find that the SBO radiation has a shallower spectral slope at low energies than blackbody radiation, meaning that the SBO optical and IR luminosities we present in \S\ref{sec:observing_sbo} assuming a blackbody spectrum should be taken as lower limits.  Nonetheless, these should be regarded as only order of magnitude estimates of the SBO flux.

\subsection{Failed SNe}
\label{sec:failed_sne}
There appears to be a paucity of higher mass progenitors to ccSNe
\citep{kochanek08,smartt09a,eldridge13}.
In particular, \cite{smartt09a} note that the maximum zero-age main-sequence mass
that seems to be associated with Type IIP SNe originating from
red supergiants appears to be $\sim 17M_\odot$, 
while stars are expected to explode as red supergiants up to
$\sim 25 M_\odot$.  Simulations show that stars in this mass
range have density structures that make it more difficult for these 
stars to explode as SNe \citep{oconnor11,ugliano12}.
While there has always been some parameter range where black
hole formation without an SNe was expected \citep[e.g.,][]{heger03},
these recent results suggest that the phenomenon is more common
than the earlier view that it would be restricted to very high
mass stars in low metallicity galaxies.  The upper limit on the fraction of ccSN that fail to explode is $\sim50\%$ with a nominal estimate of $\sim10\%$ \citep{lien10,horiuchi11}.

It is perfectly feasible to search for such failed SN based on
electromagnetic signatures because in the final analysis a
massive, luminous star effectively vanishes, potentially with
some interim transient \citep{kochanek08}.  In the Galaxy
one has the added advantage that neutrinos will clearly indicate
that a core collapse has occurred.  \cite{nadezhin80} pointed
out that the envelopes of red supergiants in this mass range are so tenuously
bound that the drop in gravitational potential energy due to
the binding energy carried by the escaping neutrinos 
during core collapse is sufficient to unbind
the envelope.  Recently, \cite{lovegrove13} have
simulated the process using realistic models of $15$ and $25M_\odot$
red supergiants and confirmed the effect.  The external signature
consists of an SBO, followed by a roughly
year long transient with a luminosity of $\sim 10^6 L_\odot$ and
an apparent temperature of order $3000$~K as the envelope
expands, cools, and releases the energy associated with
recombination.  Because the shock velocities of $\sim 10^2$~km/s
are much lower than for a true SN, the shock breakout pulse is
both much weaker and much cooler.  The \cite{lovegrove13} 
simulations show a peak of order $10^6 L_\odot$ but may not
adequately resolve the thin surface layer.  \cite{piro13} applied
analytic models of shock breakouts, finding a peak luminosity
of order $3 \times 10^7 L_\odot$ with a temperature of order
$10^4 K$ and lasting $\sim 10$~days, and determine that the shock breakout spectrum is thermal.  As a rough guide to
the detectability of such transients we make the shock breakout a $10^7 L_\odot$, $10^4$~K blackbody, and the transient
a $10^6 L_\odot$, $3000$~K blackbody.

\section{Results}
\label{sec:results}
Using our Galactic model we evaluate, in the following subsections, the prospects of observing the next Galactic ccSNe, its shock breakout, its progenitor, any precursor variability, and failed SNe.  Where relevant, we discuss both ccSNe and SNe Ia.  We also infer a Galactic SNe rate from historical SNe using our simulated observability.  We adopt the RJCE extinction model as our standard, and in most cases simply show the results for the other models in the figures.

\subsection{Prospects of Observing the Next Galactic SN}
\label{sec:sne}
\begin{figure}
  \centering\includegraphics[width=9.2cm, angle=0]{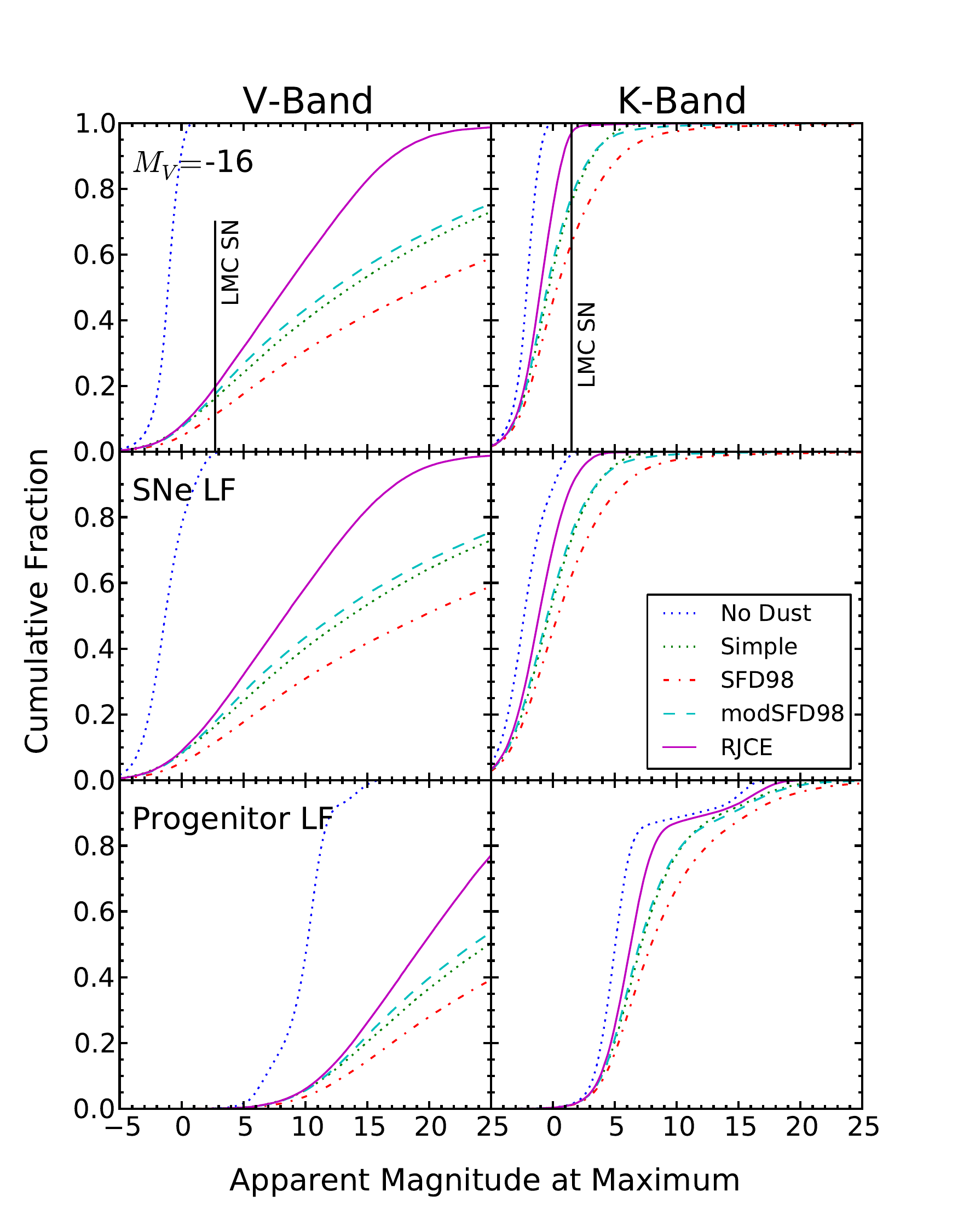}
  \caption{Cumulative magnitude probability distributions for ccSNe and their progenitors.  The top panels assume a fixed $M_{V,max} = -16$.  The middle panels use the luminosity function found by \cite{li11a}.  In both cases we use a fixed $V-K = 1.0$.  The bottom panels are derived from the Padova isochrones \citep{marigo08}.  The models for the different dust normalizations are described in \S\ref{sec:dust}.  To illustrate the importance of extinction, we show the brightness of a typical ccSN occurring in the LMC ($m_V \approx 2.7$, $m_K \approx 1.5$, assuming $M_{V,max}=-16$, a distance modulus of 18.5 \citep{pietrzynski13}, and an extinction of $A_{V}=0.2$.\label{fig-SNeII_mag}}
\end{figure}

\begin{figure}
  \centering\includegraphics[width=9.2cm, angle=0]{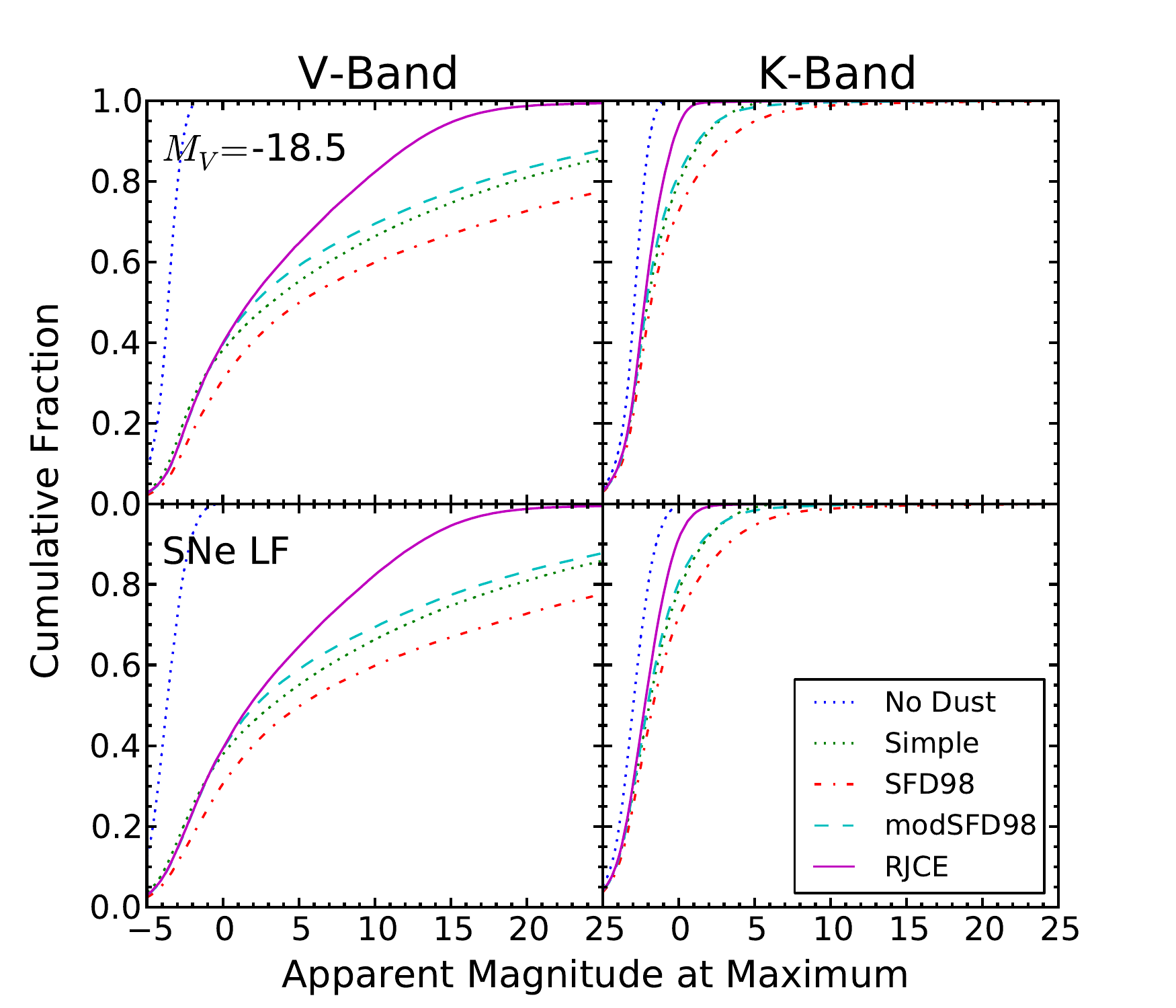}
  \caption{Cumulative magnitude probability distributions for SNe Ia.  The upper panels assume a fixed $M_{V,max} = -18.5$.  The lower panels use the luminosity function found by \cite{li11a}.  Both panels use a fixed $V-K = -0.7$ mag.  Given the model uncertainties, we make no attempt to predict the progenitor properties of SNe Ia.\label{fig-SNeIa_mag}}
\end{figure}

There is little likelihood of the next (successfully exploding) Galactic SN being unobservable.  We present the cumulative apparent magnitude distributions for ccSNe in Fig. \ref{fig-SNeII_mag} and SNe Ia in Fig. \ref{fig-SNeIa_mag}.  Most ccSNe should be observable in the optical and virtually all should be observable in the near-IR.  For example, there is a 99\% chance that the next Galactic ccSN will peak at $m_{V,max}<25$ and a $\simeq 100\%$ chance of $m_{K,max}<14.3$.  In fact, it is likely that a Galactic ccSN would be observable by semi-professional amateurs knowing where to look, with 82\% of Galactic ccSNe having $m_{V}<15$.  There is approximately a one-in-three chance that a Galactic ccSN would be visible with the naked eye ($m_{V}<5$).

SNe Ia will be even easier to observe because the delayed component lies off of the Galactic plane and will be less extinguished.  While there will be no neutrino trigger or pointing information to search for a Galactic SN Ia, 92\% will have $m_{V,max}<13.5$, which is within the limits of current all sky surveys, and, if the Large Synoptic Survey Telescope (LSST) monitors the Galactic plane, over 99\% of SN Ia would be detected.   
Confusion will have little effect on the observability of Galactic SNe because the vast majority of SNe appear relatively bright in both V and K-bands.  We also note that the position of a Galactic SN Ia could easily be determined on a time scale of weeks or months using 56Ni/56Co gamma rays \citep{timmes97,horiuchi10,diehl12,ng12}.  A Galactic ccSN could likely be observed in gamma rays given some directional information \citep{timmes97,diehl12}.

As emphasized in \S\ref{sec:ccSNe}, the magnitude distribution is primarily controlled by extinction rather than by the small spread in distance modulus across the Galaxy.  
Fig. \ref{fig:dcomp} shows that the 10th-90th percentiles of the cumulative distance probability distribution range from approximately 5-15 kpc, which corresponds to a less than 2.5 magnitude spread in distance modulus.  Meanwhile, Fig. \ref{fig:Acomp} shows that the 10th-90th percentiles of the cumulative extinction probability distribution using the RJCE model differ by approximately 15 magnitudes.

If we use the modSFD98 dust model, instead of RJCE, the predicted observability decreases substantially in V-band but remains near 100\% for both ccSNe and SNe Ia in K-band.  Confusion has a noticeable effect on the V-band observability when using the modSFD98 model because this model predicts a substantial number of SNe will be faint ($20 \lesssim m_{V,max} < 25$).  With the modSFD98 model there is a $\sim 76\%$ chance the next Galactic ccSN will peak at $m_{V,max}<25$, which decreases to $\sim 72\%$ when also requiring no brighter source within 1".  Similarly, there is an 88\% chance that a Galactic SN Ia would peak at $m_{V}<25$, and an 86\% that there will also be no brighter source within 1".

\subsection{Progenitor Characteristics}
\label{sec:progenitor}
We show the cumulative magnitude probability distribution of likely ccSNe progenitors in Fig. \ref{fig-SNeII_mag}.  We do not consider SNe Ia in this section because it is not clear what mechanism (single or double degenerate) is responsible for the majority of SN Ia events and it is clear that they would be much less luminous than ccSNe progenitors.  We again emphasize, however, that the precise astrometric position available for a new Galactic SN Ia will greatly simplify attempts to characterize the progenitor.

\begin{figure}
  \includegraphics[width=9.2cm, angle=0]{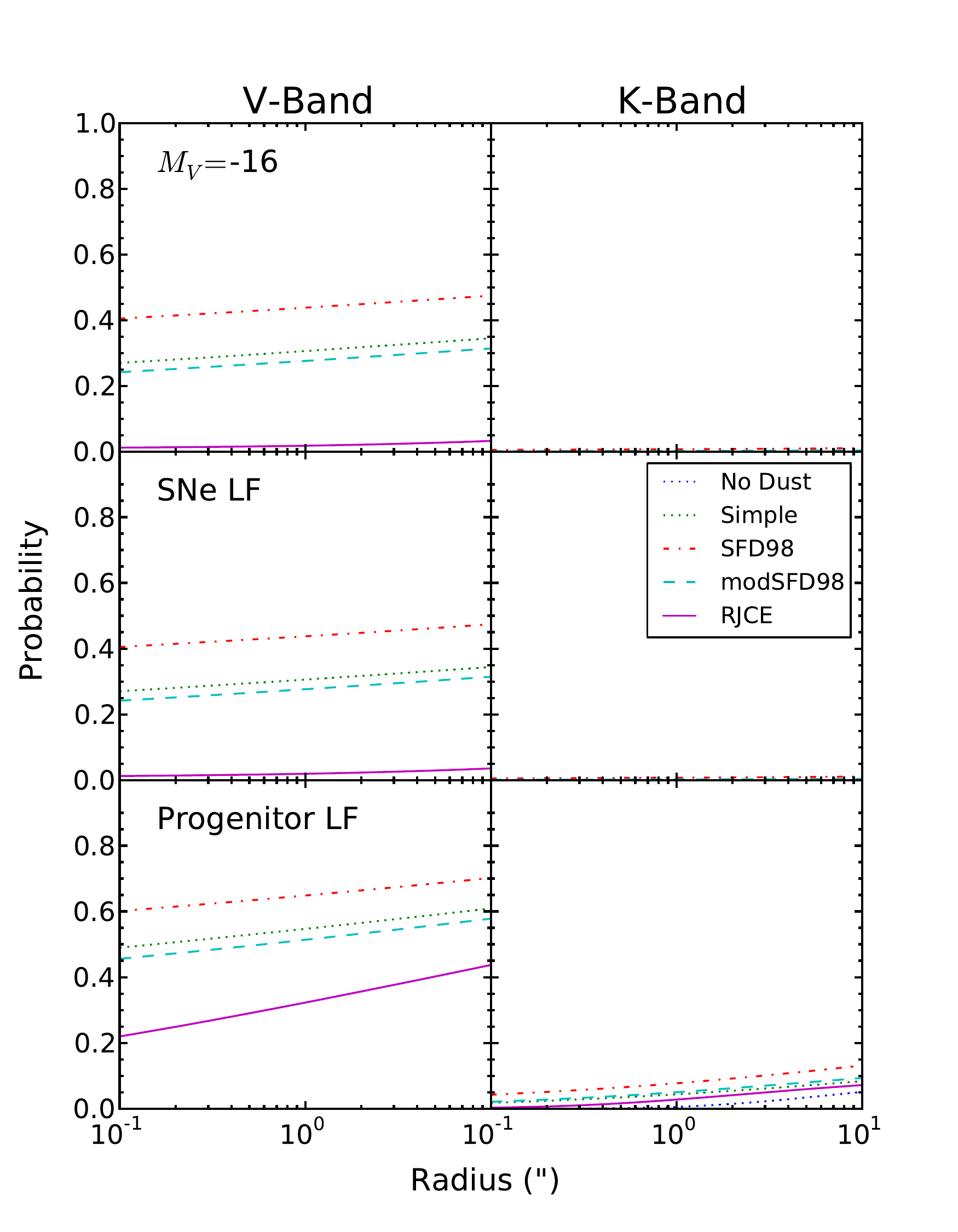}
  \caption{Probability, as a function of angular separation from a ccSNe or its progenitor, of a source with $m<m_{target}$ being present.  The panels are the same cases as in Fig. \ref{fig-SNeII_mag}.  Note that the probability of confusion affecting K-band observations of a Galactic SN is negligable.  The results for some of the dust models cannot be seen in the plot because they are essentially zero.}\label{fig-SNeIIconf}
\end{figure}

\begin{figure}
  \includegraphics[width=9.2cm, angle=0]{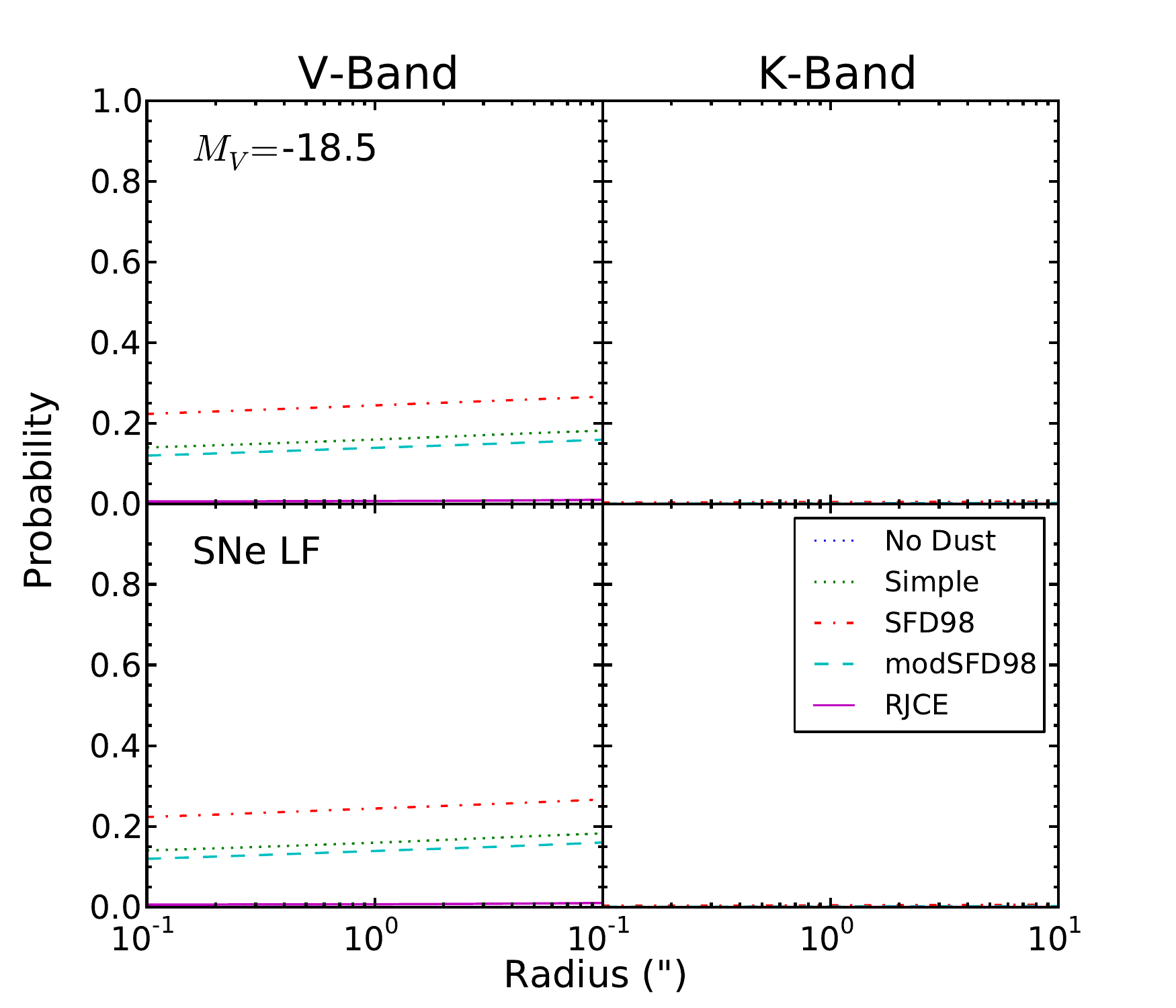}
  \caption{Probability, as a function of angular separation from a Galactic SN Ia, of a source with $m<m_{target}$ being present.  The panels are the same cases as in Fig. \ref{fig-SNeIa_mag}.  Note that the probability of confusion affecting K-band observations of a Galactic SN is negligable.  The results for some of the dust models cannot be seen in the plot because they are essentially zero.}\label{fig-SNeIaconf}
\end{figure}

Fig. \ref{fig-SNeIIconf} shows the probability of finding a star with $m<m_{prog}$ within a given radius of a ccSNe progenitor.  In the near-IR, it is unlikely that confusion is a problem.  We find that $92\%$ of likely progenitor stars have already been observed by 2MASS, assuming $m_{K,lim}=14.3$ and no brighter sources within $2"$.  
In the optical, the odds are less favorable.  We find that $57\%$ of likely progenitor stars are in the USNO-B1.0 catalog, given $m_{V,lim}=21.0$ and no brighter sources within $1"$.  A similar fraction of likely progenitors should be included in the recently completed INT Photometric H$\alpha$ Photometric Survey of the Northern Galactic Plane \citep[IPHAS;][]{drew05} and the ongoing VST/OmegaCam Photometric H$\alpha$ Survey of the Southern Galactic Plane and Bulge (VPHAS+).  A lack of optical data on the progenitor would limit our ability to physically characterize the progenitor because measurements near the peak of its spectral energy distribution are needed to constrain its temperature and thus its luminosity.  However, the ability to increase the fraction of progenitors with optical data is limited by the enormous extinction toward the Galactic center.  Even if LSST eventually images the entire sky with $m_{V,lim}=26.5$ in its coadded images, the likelihood of the progenitor being observed with no brighter sources within 1" only increases to 66\%.
The K-band results are relatively insensitive to the extinction model (with the 2MASS observability dropping slightly to 89\% with modSFD98 extinction), but the V-band results decrease significantly when using the modSFD98 model, with only 42\% of likely progenitor stars in the USNO-B1.0 catalog given the same magnitude and resolution limits, and only increasing to 48\% when considering coadded imaging from LSST.
The impact of confusion on the observability of SNe Ia will be negligible (see Fig. \ref{fig-SNeIaconf}).
We also note that difference imaging methods, scaling and subtracting post-explosion images from the pre-explosion image, can essentially eliminate confusion, as has been done for several extragalactic SNe \citep{galyam09,maund13}.  This depends on also matching the effective band passes of the data and likely cannot be applied to older photographic survey data.

Several SNe have now shown high luminosity ($10^{7}-10^{8}L_{\odot}$) eruptions in the years or months \citep{pastorello07,ofek13,mauerhan13} preceding their explosions as SNe.  While attention has focused on the dramatic but rare examples of pre-SN outbursts, there is no reason to think that the lower level of variability observed by \cite{szczygiel12} for SN2011dh is not the norm.  Few surveys exist that image the Galactic plane with the cadence necessary to detect variability in the precursor of the next Galactic SNe.  For example, the ``New Milky Way" system is being developed to survey the entire Milky Way area visible from its observing site each night to a limiting magnitude of $m_{V,lim} \sim 13.5$ \citep{sokolovsky13}.  The ASAS survey monitors the full sky to a limiting magnitude of $m_{V,lim} \sim 14$ \citep{pojmanski02} and will be extended to $\sim 16$ by the ASAS-SN survey \citep{shappee13}.  However, an all-sky catalog with $m_{V,lim} \leq 14$ (16) with no other sources brighter within 1" only includes $\sim 21\%$ (32\%) of ccSNe progenitors.  LSST has the potential to improve this situation, with 66\% of ccSNe progenitors observable with $m_{V}<24.5$ and no brighter sources within 1", however, LSST currently plans to ignore the Galactic plane \citep{abate12}.  The best method to detect precursor variability is to survey the Galactic plane in the near-IR with at least monthly cadence.  The VISTA Variables in the Via Lactea survey \citep{minniti10} will image a substantial fraction of the Galactic plane in the near-IR with sufficient cadence, but it is limited to a 5-year duration and a large fraction of likely SN progenitors will be brighter than its saturation limit.  A limiting magnitude of $m_{K}\sim 8$ (10)  would be sufficient to monitor $\sim 78\%$ (87\%) of likely ccSN progenitors for precursor variability, with the caveat that near-IR variability tends to be weaker than optical variability.

Progenitors of failed ccSNe can be as easily identified as those of successful ccSNe if there is a transient associated with the event, either the optical signature we discuss in \S\ref{sec:failed_sne} and \S\ref{sec:observing_failed_sne} or X-ray emission from accretion of residual material onto the newly formed black hole.  Even without such additional information, the progenitor may still be identifiable by its absence in post-event imaging, essentially following the approach of \cite{kochanek08}.  The challenge is separating the vanishing of the progenitor star from all other Galactic variable sources in the search region defined by the neutrino signal.  This is likely feasible because most variable sources are ``continuously" variable, while the progenitor of a failed ccSNe can only vanish once.

\subsection{Observing the Shock Breakout}
\label{sec:observing_sbo}
Using our Galactic model, we can predict the distribution of SBO apparent magnitudes, given the radiation thermalization caveat from \S\ref{sec:sbo}, as shown in Fig. \ref{fig:sbo_mag}.  An SBO occurring in the night-time sky would likely be easily observable in both the visible (P($m_{V}<20$)$\sim 0.85$) and near-IR (P($m_{K}$$<12$)$\sim 0.92$).  Because of the high radiation temperature, the natural wavelength to search for an SBO is in the UV, which can only be done from space, as proposed by \cite{sagiv13}.  For Galactic SNe and from the Earth's surface, however, the best wavelengths are actually in the near-IR.  This is a combination of the effects of Galactic absorption and the fact that the daytime near-IR sky is darker than in the optical.  The probability of a thermalized shock breakout exceeding the brightness of the daytime near-IR sky is 60, 63, and 64\% using approximate near-IR sky brightnesses from \cite{jim11} of 6.6, 5.6, and 4.9 mag per square arcsec in J, H, and K respectively.  We present a design sketch of an instrument to detect Galactic SBO pulses even in daytime in Appendix \ref{app:One}.  While a SBO would still be unobservable if the SN appears too close to the Sun, we expect only 2\% (9\%) of SNe to occur with $20^{\circ}$ ($40^{\circ}$) of the Sun.  Even in these cases where it would be difficult or impossible to detect the SBO, the duration of the SN is long enough that there will be no trouble identifying the SN as it fades and becomes observable at night.  Given the brightness of SBO events, we present in Appendix \ref{app:Two} a system capable of detecting extragalactic shock breakouts that occur within the Local Volume.

\subsection{Identifying Failed SNe}
\label{sec:observing_failed_sne}
The prospects of identifying a star dying without a dramatic SN explosion are challenging but feasible in external galaxies \citep{kochanek08}.  Within the Galaxy it is more difficult because you have to search a huge area and the unknown distance means that you cannot associate a flux with a luminosity.  However, if failed SNe associated with red supergiants follow the models of \cite{lovegrove13} and \cite{piro13}, it would likely be possible to observe the weak shock breakout and transient associated with such an event occurring within the Galaxy provided that rough positional information is obtained from neutrino or gravitational wave experiments.  

While failed SNe arising from red supergiants and their associated shock breakouts are less luminous than traditional ccSNe, their cooler temperatures make their observability comparable.  Assuming the radiation is thermalized, the absolute magnitudes, in various filters, of a typical weak shock breakout from a failed SN are $M_{B}\sim -13.3$, $M_{V}\sim -13.5$, $M_{R}\sim -13.6$, $M_{I}\sim -13.7$, $M_{J}\sim -13.6$, $M_{H} \sim -13.7$, and $M_{K} \sim -13.6$.  The fraction of such events brighter than an arbitrary magnitude in V or K can be scaled from the top panels of Fig. \ref{fig-SNeII_mag}.  Using the RJCE extinction model this corresponds to $\sim 100\%$ of events with $m_{K}<15$ and 91\% with $m_{V}<20$, which is slightly better than the observability of the normal ccSN SBOs (see Fig. \ref{fig:sbo_mag}).  However, it is noteworthy that this improved observability translates to $\sim 97\%$ of such shock breakout events appearing brighter than the near-IR sky ($m_{K}<4.9$).
Similarly, for the transient, $M_{B}\sim -6.9$, $M_{V}\sim -8.7$, $M_{R}\sim -10.0$, $M_{I}\sim -11.2$, $M_{J}\sim -12.2$, $M_{H} \sim -12.9$, and $M_{K} \sim -13.3$, which translates to $\sim 100\%$ of events with $m_{K}<15$ and 72\% with $m_{V}<20$.

\subsection{Estimates of the Galactic SNe Rate}
We can also use the magnitude distribution of Galactic SNe from \S\ref{sec:sne} together with historical SNe to estimate the frequency of Galactic supernovae, the Galactic star formation rate (SFR), and the ratio of Galactic core-collapse to Type Ia SNe.  These estimates, however, are limited by the small number of recorded SNe and the completeness of the record.  \cite{stephenson02} conclude that 5 Galactic SNe have been observed since 1000 AD, when the historical records become relatively complete.  These supernovae are SN~1006 (SNIa), SN~1054 (ccSN), SN~1181 (ccSN), SN~1572 (SNIa), and SN~1604 (SNIa).  \cite{clark77} estimate that the apparent brightness of SN1181 was $\sim0$ mag, while the other 4 SNe were $\lesssim-4$ mag.  Given the magnitude probability distributions of our models, there is only a $\simeq 3\%$ chance that only one SN occurred with $-3<m_{V,max}<0$ if 4 occurred with $m_{V,max}<-3$.  This suggests that the historical record may be incomplete for SNe fainter than $m_{V,max}\lesssim-2$.  Therefore we separately present results using the $m_{V,max}<-2$ (3 SNeIa and 1 ccSNe) and the full SN samples (3 SNeIa and 2 ccSNe), but take the results found using the $m_{V,max}<-2$ sample to be more meaningful.  While the supernova that produced Cas A might have been observed in 1680 as a 6th magnitude event \citep[see, e.g.,][]{thorstensen01}, we do not consider it in our analysis since the historical record is clearly incomplete for such faint SNe.

Our results from \S\ref{sec:sne} show that $3.6\%$ ($9.0\%$) of Galactic ccSNe and $24\%$ ($40\%$) of SNe Ia have $m_{V,max} < -2$ (0).  SNe Ia are more observable than ccSNe both because they are intrinsically brighter and because the delayed SNe Ia component suffers from far less extinction due to its larger scale height.
Historical SNe were recorded almost exclusively by cultures in the Northern hemisphere, with China providing the most complete record.  
Since 1000 AD, the Chinese capitals, where the observations that are the basis of the historical records were made, were located primarily between $30^{\circ}$ and $40^{\circ}$ N.  We find that for a fiducial latitude of $35^{\circ}$ N, $90\%$ of Galactic SNe with $m_{V,max}<-2$ would have been above the horizon at night during their peak.

The number of historical SNe, folded together with our simulated observability at $35^{\circ}$ N, suggests a Galactic core-collapse SN rate of $3.2^{+7.3}_{-2.6}$ ($2.5^{+3.4}_{-1.6}$) per century and a Galactic Type Ia SN rate of $1.4^{+1.4}_{-0.8}$ ($0.8^{+0.8}_{-0.5}$) per century for a total Galactic SN rate of $4.6^{+7.4}_{-2.7}$ ($3.4^{+3.4}_{-1.7}$) per century using $m_{V,max}<-2$ (0) limits.
Repeating this exercise using the modSFD98 extinction (instead of RJCE), we find that $3.4\%$ ($8.2\%$) of Galactic ccSNe and $24\%$ ($39\%$) of SNe Ia have $m_{V,max} < -2$ (0).  This corresponds to $7.2\%$ ($15\%$) of all Galactic SNe having $m_{V,max} < -2$ (0), a Galactic core-collapse SN rate of $3.4^{+7.8}_{-2.8}$ ($2.8^{+3.7}_{-1.8}$) per century, and a Galactic Type Ia SN rate of $1.4^{+1.4}_{-0.8}$ ($0.8^{+0.8}_{-0.5}$) per century for a total Galactic SN rate of $4.8^{+7.9}_{-2.9}$ ($3.7^{+3.8}_{-1.9}$) per century using $m_{V,max}<-2$ (0) limits.  The different extinction models we test give fairly consistent results for the inferred Galactic SN rate because these bright SNe must be relatively nearby where the details of the dust model are relatively unimportant.
The SN rates we infer are in reasonable agreement with other estimates that are found by a variety of methods, including historical Galactic SNe \citep[][with $2.5^{+0.8}_{-0.5}$ SN/century and $5.7\pm1.7$ SN/century respectively]{tammann94,strom94}, the massive star birthrate \citep[][1-2 ccSN/century]{reed05}, radioactive aluminum from massive stars \citep[][$1.9\pm1.1$ ccSN/century]{diehl06}, the pulsar birthrate \citep[][$2.8\pm0.1$ ccSN/century and $10.8^{+7}_{-5}$ ccSN/century respectively]{keane08,faucher06}, and the extragalactic SN rate by Hubble type and stellar mass \citep[][$2.8\pm0.6$ SN/century]{li11b}.

We use the SFR to core-collapse SN rate conversion coefficient of $0.0088/\mathrm{M}_{\odot}$ from \cite{horiuchi11}, which assumes a modified Salpeter initial mass function, to infer a Galactic SFR from our calculated rate of Galactic ccSNe.  Using the rate based on the RJCE extinction model and the $m_{V,max}<-2$ ($m_{V,max}<0$) sample of ccSNe we estimate the Milky Way's SFR to be $3.6^{+8.3}_{-3.0}$ ($2.9^{+3.8}_{-1.9}$) M$_{\odot}$ yr$^{-1}$, where the quoted uncertainties are purely statistical.  This SFR is consistent with direct estimates of the SFR, which range from 1 to 4 M$_{\odot}$ yr$^{-1}$ \citep[e.g.,][]{mckee97,murray10,robitaille10,chomiuk11,davies11}.

We can also use the observed fraction of each type of SN to place weak limits on the total fraction of each type of SN in the Milky Way.  One of the four (2 of 5) observed Galactic SNe with $m_{V}<-2$ (0) since 1000 AD were ccSNe, which, folded together with the relative observability of core-collapse and Type Ia SNe, suggests that the fraction of Galactic SNe that are ccSNe, $f_{ccSN}$, is $0.69^{+0.22}_{-0.46}$ ($0.75^{+0.16}_{-0.31}$).  This result is consistent with the $f_{ccSN}=0.81$ found for Milky Way-like galaxies by \cite{li11b}.

\begin{figure}
  \includegraphics[width=9.2cm, angle=0]{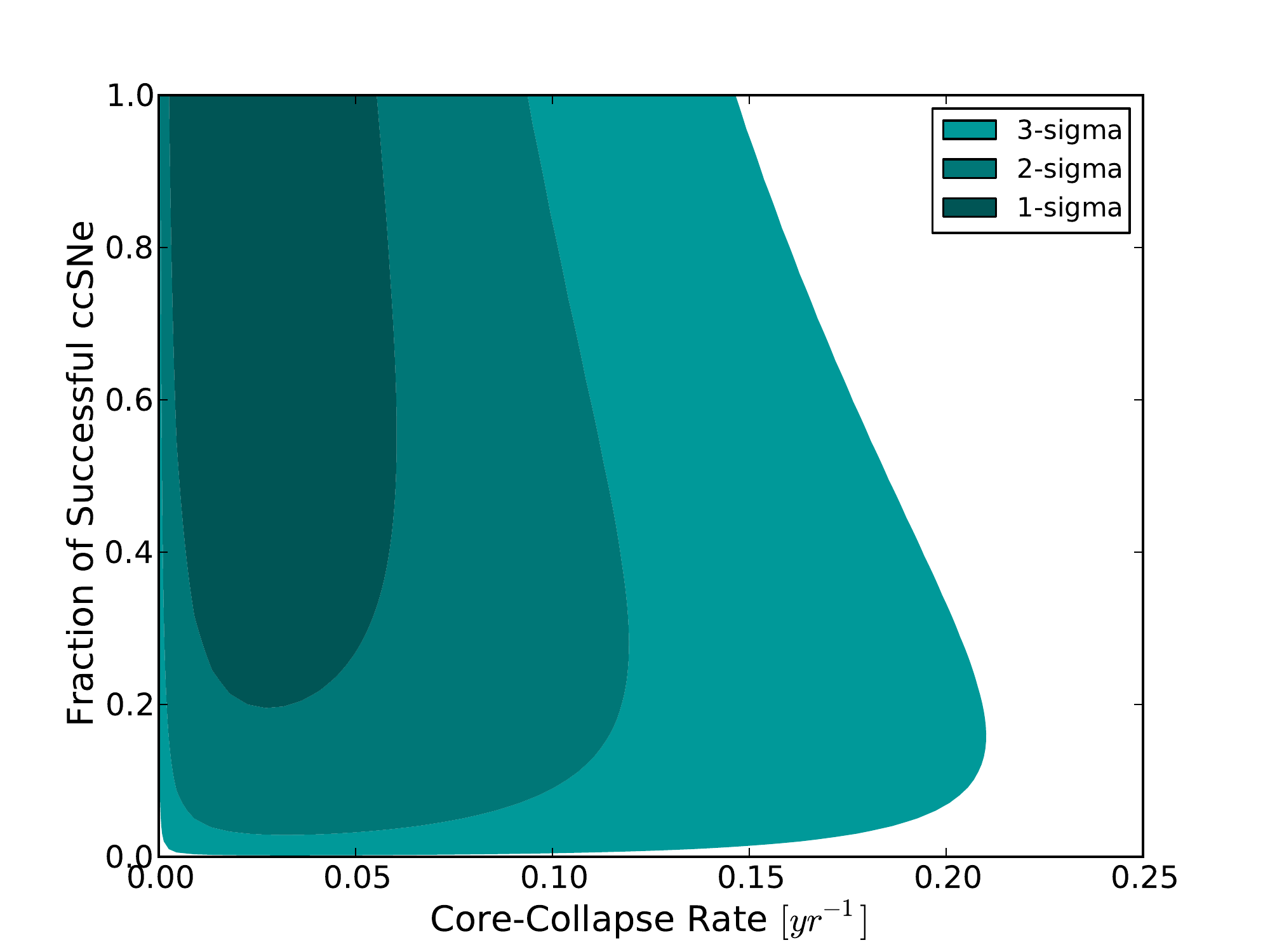}
  \caption{Limits on the fraction of core-collapse events that produce normal, luminous ccSNe as a function of the total Galactic core-collapse rate.  These limits are found by combining the rate of Galactic ccSNe we infer from the historical record with the upper limits placed on the rate of Galactic core-collapse events by the non-detection of SN neutrinos by 30 years of active neutrino detection experiments.
\label{fig-failed_vs_Rcc}}
\end{figure}

The SN rate, SFR, and fraction of SN by type inferred from our model all have large uncertainties due to the limited number of historical SN, but they demonstrate that the predicted observability of Galactic SNe given by our models is reasonable.  While consistent with estimates of the Galactic SN and star formation rates due to the large uncertainties, the rates are somewhat high, which could suggest that the SN rate within a few kpc of the Earth is higher than the Galactic mean.
We note that the existence of ``failed" SNe would increase our inferred Galactic core collapse and star formation rates.  
Combining the upper limit placed by the non-detection of SN neutrinos over $\sim30$ years of measurements by neutrino detection experiments \citep{alexeyev02,ikeda07} of $\lesssim 8$ core-collapses/century with our inferred luminous ccSN rate allows us to place weak limits on both the rate of core-collapse events and the fraction of such events that fail to produce normal, luminous ccSNe (see Fig. \ref{fig-failed_vs_Rcc}).  
This limit on the fraction of core-collapse events that fail to explode is consistent with the upper limit of $\sim 50\%$ found by \cite{horiuchi11} and the nominal estimate of $\sim 10\%$ given by \cite{lien10}.

\section{Importance of Neutrino Detection}

As noted in the introduction, the detection of a burst of MeV neutrinos can provide crucial answers to three questions: \emph{IF astronomers should look for a Milky Way supernova}, \emph{WHEN they should look}, and \emph{WHERE they should look}.

In the following, we review the role of neutrinos in understanding collapses, the basics of their production and detection, and the present state of inter-experimental co-operation.  We then provide new information on the alert procedure of the Super--Kamiokande Collaboration.  Most importantly, we also provide the first announcement of a new fast-response capability using the EGADS experiment.

The case of SN 1987A is instructive for defining the main issues \citep{arnett89}.  The neutrinos were detected a few hours before the deduced start of the electromagnetic signals, though this was not realized until later.  At the time, the world first became aware of the event through optical detections of the early light curve \citep{kunkel87}, which were fortuitous and might have been missed until the supernova became brighter.  The neutrinos were easily detected even though the detectors of that time were relatively small and the Large Magellanic Cloud is about five times farther than a typical Milky Way supernova (see Fig. \ref{fig:dcomp}).  The progenitor star was detected in archival images \citep{walborn87}, and little information was available on possible variations in its pre-explosion luminosity \citep[see][]{plotkin04}.  The SBO and the earliest supernova light curve were also undetected.

For the next Milky Way supernova, neutrinos should serve as the starting gun indicating that the race is on to characterize this once-in-a-generation event in detail across many timescales and electromagnetic bands.  The key advantage of neutrinos, besides answering the three questions above and thus possibly allowing one last look at the undisturbed star just before it is destroyed, is that they can reveal the conditions and dynamics deep within the star.  A primary goal that we emphasize is to catch not just the early supernova light curve, but also the SBO that precedes it.  This will require getting alerts, times, and directions from the neutrino experiments far more rapidly than envisaged by the current system.  Some of the present detectors are considerably larger and all have much swifter data processing than those in 1987, but existing data-sharing plans may still lead to crucial lost opportunities.

\subsection{Neutrino Production in Core Collapse}

The protoneutron star formed after core collapse is nearly at nuclear density and is at a central temperature of tens of MeV.  It sheds almost all of its energy by radiating neutrinos, mostly through neutrino pair-production processes (the neutronization process $p + e^- \rightarrow n + \nu_e$ corresponds to only $\sim 10\%$ of the total neutrino emission).  Because the density is so high, even neutrinos have difficulty escaping from the proto-neutron star, and they diffuse out over several seconds with a spectrum characteristic of the temperature, $T$, at the surface of last scattering, typically a few MeV.  There are thought to be differences between the six neutrino flavors ($\nu_e$, $\nu_\mu$, $\nu_\tau$, and their antiparticles) in terms of their total and average energies, but it is a reasonable simplification to say that each flavor should carry about 1/6 of the binding energy release of $\sim (3/5) G M^2 / R \sim 3 \times 10^{53}$ erg and have an average energy of $\simeq 3T \simeq 15$ MeV. 

Core collapses are extremely efficient neutrino generators and the neutrinos carry about $10^4$ times more energy than the eventual optical supernova.  All smaller, less efficient explosions cannot possibly produce enough neutrinos to be reliably detected.  The neutrino emission from a Type Ia supernova arises just from accelerated nuclear burning, is much more modest, and does not include the most detectable flavor, $\bar{\nu}_e$.  Supernova impostors, if they really are just non-destructive outbursts of massive stars, should produce essentially no neutrinos.  For collapses that lead to black hole production and little or no electromagnetic emission, the time-integrated neutrino signals are similar, essentially because the thermal energy of the hot proto-neutron star must be lost before the final collapse \citep{nakazato07,oconnor11}.  Black-hole forming events will show a distinctive truncation of the neutrino signal in time \citep{beacom01,nakazato12}, which would be very relevant for the subsequent electromagnetic searches.

\subsection{Detecting Supernova Neutrinos}
\label{sec:neutrino_detection}

The most important supernova neutrino detection reaction is inverse beta decay, $\bar{\nu}_e + p \rightarrow e^+ + n$, where the proton is a hydrogen nucleus \citep[see][]{scholberg12}.  The total energy of the outgoing positron is $E_e \simeq E_\nu - 1.3 {\rm\ MeV}$, and its direction is nearly isotropic because of the small recoil energy of the nucleon.  There are also interactions with electrons and with nucleons bound in nuclei, but they have smaller cross sections and less favorable kinematics for the detectable particles in the final states. 
The Super--Kamiokande detector \citep{fukuda03} in Japan has a nominal fiducial volume of 22.5 kton of ultra-pure water, and for a core collapse at the Milky Way center, Super--Kamiokande expects to detect $N_{\nu p} \sim 10^4$ inverse beta events over several seconds, with a negligible number of background events.   The reconstruction directions of these neutrino events will be nearly isotropic \citep{vogel99}. Super--Kamiokande will have excellent measurements of the $\bar{\nu}_e$ energy spectrum and luminosity profile, as well as information on other neutrino flavors using other detection reactions.  From the number of events, the distance to the supernova could in principle be estimated with percent-level precision; however, uncertainties in the emission models will likely restrict this to a few tens of percent.

The IceCube detector is vastly larger than Super--Kamiokande but has a very high detector background rate, which means that individual neutrino interactions cannot be separated from non-neutrino events.  Nevertheless, for a core collapse in the Milky Way, the number of neutrino interactions would be so large in such a short time that IceCube would see a highly significant increase in the apparent ``background" rate, yielding an unambiguous detection \citep{abbasi11}.  IceCube will have excellent data on the luminosity profile but no information on the energies or flavors of individual events.  There are various other smaller detectors that will also provide important confirmations of a neutrino burst and some additional information about the other neutrino flavors \citep{scholberg12}.  The range of existing detectors is just the Milky Way and its immediate companions, from which a burst could be detected easily, and not any nearby galaxies.  For a core collapse in Andromeda, Super--Kamiokande would detect $\sim 1$ event and other detectors would detect nothing; much larger detectors will be needed to probe the much-more frequent extragalactic events \citep[][but see also Appendix \ref{app:Two}]{ando05,kistler11}.

Another important detection reaction in Super--Kamiokande is neutrino-electron scattering, for which the cross section is smaller.  Summed over all flavors, hundreds of events are expected.  Because the electron mass is small compared to the neutrino energies, the electrons are scattered forward, within $\sim 10^\circ$ of the neutrino direction.  Taking into account various systematics, Super--Kamiokande should be able to constrain the direction to a Galactic ccSN to within a few degrees \citep{beacom99}.  The prospects for directionality from timing triangulation using multiple detectors are poor, due to the long timescales and low statistics relative to the Earth-crossing time \citep{beacom99}.

\subsection{Neutrino Alert of Core Collapse}

As the neutrinos are generated and arrive at Earth before any electromagnetic signals, there will be a brief period, hours at most, during which the detected neutrinos are the only indication, other than gravitational waves, of an ongoing core collapse event.  This places a high burden of responsibility on the neutrino experiments to announce as much information as soon as possible.

On the other hand, Milky Way core collapses are so rare that false signals, perhaps related to detector electronics, may occur during the decades-long waits.  This gives a strong motivation to the experimentalists to be very careful to avoid announcing possibly false signals.  The human intervention required to have adequate confidence may take hours.

There lies peril on both sides: the consequences of failing to react swiftly to a true signal or of raising a false alarm could be quite detrimental to the reputation of the neutrino experiments.   Of these two unpleasant scenarios, however, avoiding even the possibility 
of issuing a false alarm has traditionally been of greater concern to experimentalists.  None of the present detectors, with the possible exception of Baksan \citep{alekseev87}, has ever seen a core-collapse neutrino burst.  Couple this with the difficulty inherent in devising a calibration method with which to properly mimic uniformly volume-distributed, SN-like light in the detectors, and the problem becomes evident: how can an experimental collaboration rapidly muster sufficient confidence that what they are seeing is, in fact, a supernova, and get the word out to astronomers in time to catch the SBO?

\subsubsection{SNEWS}

In an early attempt to get around the problem of false alarms and to get word out quickly, a few of the world's supernova neutrino detectors have banded  together via the SuperNova Early Warning System, SNEWS \citep{antonioli04}.  The essential idea is simple: if a  participating detector believes it is seeing a supernova neutrino burst, it sends the time of the start of the burst (and any other data it wishes to release) to a central server.  If several geographically separated detectors report a burst within a certain period then an automated alert is sent to a pre-determined, open subscription email list.  SNEWS has been operating in one configuration or another since 1998 with between two and four detectors in the network.  It currently contains Super--Kamiokande, IceCube, Borexino, and LVD\footnote{\url{http://snews.bnl.gov/}}.

In order to avoid false alarms and potential embarrassment of the experimental collaborations, by binding agreement, no information whatsoever is ever released unless the SNEWS coincidence threshold determined  by the number of participating detectors is reached. In addition, the false alarm rate of individual participating detectors is required to be low enough --- roughly once per month --- that random false coincidences should occur less than once per century.  Detectors which for any reason temporarily exceed this agreed upon rate are excluded during that period from participating in coincidences.

Unfortunately, the statistics are such that based upon arrival times and size of the burst alone --- which is all the participating experiments have agreed to release prior to human review --- the direction of the burst cannot be located \citep{beacom99}, relegating SNEWS to serve as a wakeup call for those who have signed up.  Furthermore, given the limited number of participating collaborations, their overriding desire to avoid false positives, and the resulting strict coincidence conditions, there is inevitably some risk of a false SNEWS negative.  Ironically, the magnitude of this important risk cannot be known due to the nondisclosure agreements regarding individual detector performance (i.e., uptime fraction and false alarm rate) that made SNEWS possible in the first place.

\subsubsection{SN Procedure in Super--Kamiokande}

While it is critical to get the directional information out as soon as possible, only Super--Kamiokande will have quality pointing data --- a few-degree error circle --- for a burst anywhere in the Milky Way.  Therefore we must consider that crucial experiment's (previously unpublished) approach to supernova data review and release.  Within the Super--Kamiokande collaboration exists a standing committee called SURGE, the Supernova Urgent Response Group of Experts.  If Super--Kamiokande's near-realtime data analysis processes identify a sudden burst of supernova neutrino-like events in the detector, within approximately two minutes an alert containing the time of the burst is sent to SNEWS.

These processes also initiate a specific, pre-arranged, and rehearsed procedure.  Automated phone calls are placed to the SURGE members as well as the experiment's Spokesman and other executive committee members, about twenty people in all.  Burst data in various forms are also sent to their mobile phones for review.  A video conference is convened within 15 minutes of the burst's detection, during which the operating condition of the detector is verified, key plots are discussed, and characteristic event displays are shown.
 
If it is agreed that a real supernova has been detected, the rest of the Collaboration is notified via email and pre-worded  announcements are sent to IAU and ATel.  These telegrams contain the universal time of the start of the burst, its duration, how many neutrinos above 7 MeV were observed during that time, the right ascension and declination of the radiant, and the errors on the fitted direction.  This is all designed to take place in an hour or less.  Drills are held to work out sticking points in the procedure and speed up the entire process as much as possible.  Nevertheless, it is clear that the experiment's very careful, very responsible, hands-on approach to building the locally required level of confidence and consensus means that Super--Kamiokande cannot issue its supernova alert as quickly as would be ideal.

\subsubsection{EGADS and Instant Alerts}

How can we build the confidence needed to get the announcement response time of even a single neutrino detector below the one-minute mark, while simultaneously minimizing the risk of both false positives and false negatives?  

Presently, the Super--Kamiokande detector can only detect the positrons in the inverse beta decay reaction, $\bar{\nu}_{e} + p \rightarrow e^{+} + n$.  If we could detect the neutrons in coincidence, then we could greatly increase the certainty that a supernova was occurring, as there are very few physics processes that could mimic this coincident signal, and none that could fake a burst of many such events.
As was first pointed out some years ago \citep{Beacom04}, adding a 0.2\% solution of a water-soluble gadolinium compound like gadolinium chloride  or gadolinium sulfate to light water Cherenkov detectors would allow such coincident detection, within displacements of centimeters in space and tens of microseconds in time.  Gadolinium has a thermal neutron capture cross section of 49,000 barns (about 5 orders of magnitude larger than that of protons) and emits a gamma cascade of 8 MeV that can be easily detected by detectors like Super--Kamiokande.  This assertion has since been verified in the Super--Kamiokande detector itself via the use of a gadolinium-loaded calibration source \citep{watanabe09}.

In an effort to prove that a gadolinium-enriched Super--Kamiokande will be effective,
 starting in 2009 a large-scale test facility called EGADS, Evaluating Gadolinium's Action on Detector Systems, was built in the Kamioka mine \citep{vagins11}\footnote{\url{http://www.ipmu.jp/webfm_send/555}}$^{,}$\footnote{\url{http://hanse2011.desy.de/sites/site_hanse2011/content/e119287/e119757/Vagins_HANSE11.pdf}}.  
EGADS's centerpiece is a 200-ton water tank, essentially a $\sim 1\%$ scale model of Super--Kamiokande, complete with a novel, selective water filtration system \citep{vagins12} and 240 50-cm photomultiplier tubes.  
The gadolinium studies have gone well, and this facility will soon be repurposed for an exciting new (and previously unpublished) role.

As part of a new multimessenger supernova astronomy initiative in Japan\footnote{\url{http://www.gw.hep.osaka-cu.ac.jp/gwastro/A03/overview_e.html}}, in 2014 EGADS will be converted --- primarily via upgraded DAQ electronics, addition of computing sufficient for 100\% real-time event reconstruction, and improved calibration --- from an R\&D testbed to a dedicated supernova neutrino detector.  As part of this process, the EGADS acronym will be redefined to mean ``Employing Gadolinium to Autonomously Detect Supernovas".  For a core-collapse event near the Milky Way center, $\sim 100$ events will be detected.

This modestly-sized detector will become an especially important supernova neutrino detector, due to a unique and vital capability: bursts will be detected and  announced to the world within {\em one second} of the first neutrino's arrival in the tank.  The confidence required to do this is provided by the gadolinium loading.  If the ``heartbeat'' signature of several coincident inverse beta decay events is seen, the double flash of positron Cherenkov light quickly followed by neutron capture gammas from the same spot in the  detector, then a Milky Way burst is most assuredly under way and can be announced immediately without human intervention.

This data will have no directionality unless there is an especially close event (EGADS would see $\sim$100,000 events from Betelgeuse).  While officially a standalone, independently-funded project distinct from Super--Kamiokande, all the members of the considerably smaller EGADS Collaboration are in fact members of both collaborations.  What is more, the PI of the  R\&D-phase EGADS (Masayuki Nakahata) and the PI of the supernova detector-phase EGADS  (MRV, one of this paper's authors) are both members of SURGE and are the co-conveners of Super--Kamiokande's solar and supernova neutrino analysis group. It is therefore hoped that an immediate, positive supernova detection in EGADS will supply the neighboring Super--Kamiokande, even if it has not yet been enriched with gadolinium itself, with sufficient confidence to react much more quickly in releasing its critical directional information.
With EGADS, or a similar solution to the problem of timeliness, and a modest investment as outlined in Appendix \ref{app:One}, there is no fundamental barrier to observing even the short ephemeral SBO signature of a Galactic ccSN.

\section{Summary and Conclusions}
The scientific community is eagerly awaiting the next Galactic SN.  A Galactic ccSN will allow the application of an array of probes that are not possible for the many extragalactic SNe that are observed.  Using modern dust models we provide a detailed assessment of the observability of the next Galactic SN, including, for the first time, near-IR estimates, the effect of confusion, and the observability of ccSN progenitors, precursors, shock breakouts, and failed SNe.

We find that a Galactic ccSN (assuming a successful explosion) will be observable in the near-IR ($P(m_{K}<5)\simeq 100\%$) and very likely ($P(m_{V}<20)\sim96\%$) will be observable in V-band.  A Galactic ccSN will produce an unmistakable neutrino signal to easily trigger electromagnetic searches.  For a Milky Way SN, the Super--Kamiokande detector will localize the SN position to within a few degrees.  Given the ccSN magnitude probability distribution we find, along with the expected neutrino pointing uncertainty, it will be possible for wide field near-IR and optical instruments to identify the SN.

While $\sim4/5$ Galactic SN are of the core-collapse variety \citep{li11b}, we note that a Galactic SN Ia is likely to be appear brighter than a ccSN because SNe Ia are intrinsically brighter than ccSNe and their spatial distribution has a larger scale height, which results in less average line-of-sight extinction.  Although a Galactic SN Ia would not produce a (currently) detectable neutrino signal to trigger a search, we find that 92\% will have $m_{V,max}<13.5$, which is within the limits of current all sky surveys, and if the Large Synoptic Survey Telescope (LSST) monitors the Galactic plane over 99\% of SN Ia would be detected.

We use our modeled observability of Galactic SNe, together with the record of historical SNe, to estimate the rate of Galactic SNe, the ratio of Galactic core-collapse to Type Ia SNe, and the Galactic SFR.  We infer a Galactic ccSN rate of $3.2^{+7.3}_{-2.6}$ per century and a Galactic SN Ia rate of $1.4^{+1.4}_{-0.8}$ per century for a total Galactic SN rate of $4.6^{+7.4}_{-2.7}$ per century and constrain the fraction of Galactic SNe that are ccSNe to be $0.69^{+0.22}_{-0.46}$.  We in turn use this Galactic ccSN rate to infer a Galactic SFR of $3.6^{+8.3}_{-3.0}$ M$_{\odot}$yr$^{-1}$.

Combining the upper limit placed by the non-detection of SN neutrinos over $\sim30$ years of measurements by neutrino detection experiments \citep{alexeyev02,ikeda07} of $\lesssim 8$ core-collapses/century with our inferred luminous ccSN rate allows us to place weak limits on both the rate of core-collapse events and the fraction of such events that fail to produce normal, luminous ccSNe (see Fig. \ref{fig-failed_vs_Rcc}).
This limit on the fraction of core-collapse events that fail to explode is consistent with the upper limit of $\sim 50\%$ found by \cite{horiuchi11} and the nominal estimate of $\sim 10\%$ given by \cite{lien10}.

We show that a Galactic ccSN could provide a unique opportunity to obtain detailed observations of the SN shock breakout.  We present expected absolute and apparent magnitude probability distributions for the SBO.  We also consider the possibility of detecting the weak SBO of a failed SN occurring when a red supergiant's envelope becomes unbound after its core undergoes a direct collapse to a black hole.  While failed SNe from other progenitors (such as blue supergiants or Wolf-Rayet stars) would not produce this sort of signal, we show that those following the red supergiant models of \cite{lovegrove13} and \cite{piro13} would be even easier to detect in the near-IR than the SBOs of normal ccSNe.  We note that given the scale of investment in neutrino detection, a dedicated day and night IR SBO detection system could be built and operated at moderate cost (see Appendix \ref{app:One}) and that the existence of such a system could further reduce the potential embarrassment of false detections---it would simply represent another layer of the overall trigger system.

Since the SBOs of normal SNe occur on timescales ranging from minutes to a couple of days after the neutrino burst, it is of paramount importance that neutrino detection experiments provide directional information in near real-time in order for an electromagnetic detection of the SBO to be possible.  We describe the procedure Super--Kamiokande will follow in the event of a SN detection before releasing this information.  We further describe EGADS, a system that will provide instant Galactic SN alerts.  We also present an outline for a system capable of detecting extragalactic SN SBOs (see Appendix \ref{app:Two}) that could better cue searches for neutrino bursts that could not be detected directly.

We also model the observability of likely ccSN progenitors.  We find that $\sim 92\%$ already have near-IR imaging (with 2MASS), but only $\sim 57\%$ have V-band imaging (in the USNO-B1.0 catalog).  A lack of optical imaging of the progenitor would limit our ability to characterize its temperature and luminosity.  However, the enormous extinction towards the Galactic center will make it difficult to substantially increase the fraction of likely progenitors imaged in V-band.  We also consider the potential for observing precursor outbursts, but find that current all-sky surveys are not likely to observe such events.  We note that LSST (if it images the Galactic plane) or a shallow near-IR all-sky survey would be capable of monitoring a majority of likely progenitors for precursor variability.  Unfortunately, LSST presently plans to largely ignore the Galactic plane, which, as recently discussed by \cite{gould13}, may be a suboptimal strategy.

This paper shows that the astronomical community could make important observations of the stages leading up to and including the traditional SN light curve of the next Galactic SN.  
However, there are steps that must be taken by astronomers and neutrino experimentalists working together to insure that the next rare opportunity to observe a Galactic SN is not squandered.

\begin{acknowledgements}
We thank Todd Thompson, Rick Pogge, Bruce Atwood, Jos\'{e} Prieto, and Ben Shappee for valuable discussions and information.  We thank Elizabeth Lovegrove, Stan Woosley, and Tony Piro for advanced looks at their relevant works on modeling the explosions and shock breakouts of failed SNe.  We thank Georg Raffelt and the anonymous referee for helpful comments.  J.F.B. is supported by NSF grant PHY-1101216.  M.R.V. is supported by the World Premier International Research Center
Initiative of Japan's Ministry of Education, Culture, Sports, Science, and
Technology (MEXT), and also by a MEXT Grant-in-Aid for Scientific Research
on Innovative Areas (24103004).
\end{acknowledgements}

\begin{appendix}
\section{Instrumentation for Shock Breakout Detection in Daytime}
\label{app:One}

Here we outline a rough design for an instrument capable of detecting Galactic SN shock breakout bursts even in daytime.  This particular design minimizes cost at the price of sensitivity.  Sensitivity can be rapidly gained using a larger aperture at the cost of a smaller field of view or more detectors.  While the shock breakout pulse peaks in the far-UV/X-ray, a search for a Galactic breakout pulse should be done in the near-IR because of Galactic extinction and the brightness of the daytime sky.  As noted earlier, we are making assumptions about the thermalization of the SBO radiation, but based on \cite{sapir13} they are conservative.

We consider an 86~mm aperture, 300~mm focal length, short
wavelength IR (SWIR) lens with a 20.5~mm focal plane
corresponding to a 3.9~degree field of view (an Optec
OB-SWIR300/3.5 SWIR lens).  We then use a FLIR Systems
InGaAs $640\times 512$ detector (a Tau SWIR BP detector)
with 25$\mu$m pixels.  This provides a $3.0 \times 2.5$~degree
field of view with $17$~arcsec pixels that is well-matched
to the positional uncertainties expected from Super--Kamiokande (see \S\ref{sec:neutrino_detection}).  The detector QE (80\%)
and optical through-puts (50\%) are not grossly dissimilar
from those of an astronomical infrared instrument, but the
read noise and well depths are much higher ($400$ and
$2.5 \times 10^6$~e$^-$, respectively), and the detector
can be read with a frame time of 30~Hz.  If we scale from
the Lucifer exposure time calculator\footnote{\url{http://www.lsw.uni-heidelberg.de/lucifer-cgi/calculator/calculator.py}} for the Large Binocular
Telescope \citep{ageorges10}, we expect 7500 and 18000 counts/second for J/H=6~mag.
However, with the large pixels, the sky count rates are
$2.2$ and $5.5\times 10^6$/second, respectively, so to avoid
saturation one would operate near the maximum frame rate
\footnote{Daytime IR sky brightnesses are approximately 6.6 and 5.6 mag per square arcsec in J and H respectively \citep{jim11}, while the daytime V-band sky brightness is $\sim 4$ mag per square arcsec \citep{rork82}.}.
For these nominal parameters, the signal-to-noise ratio 
is roughly $3$ or $6 \times t^{1/2}10^{-0.4(m-6)}$, for a
source with a J or H magnitude of $m$ and an integration time of $t$ seconds.
The lens and detectors
are of moderate estimated price ($\sim\$$40,000), so it would be
entirely feasible to set up a system that would have a high probability of detecting the shock breakout pulse of the next Galactic supernova in either day or night time (unless the SN occurred at a very small Sun angle) by distributing several
such systems around the globe.  Four such facilities to cover north/south, east/west, and weather would provide a reasonably high probability of success.  The system would also supply a continuous near-IR variability survey of the Galaxy.
Note that such a system requires the directional information supplied by Super--Kamiokande to detect the SBO, although on longer time scales it could search the Galactic plane for a SN with no neutrino pointing information.

This particular design sketch is meant to minimize cost while maintaining a field of view comparable to the localization accuracies of neutrino and gravitational wave detectors (see \S\ref{sec:observing_sbo}).  Sensitivity scales with aperture, $D$, as $D^{2}$, so increasing the aperture rapidly increases the sensitivity.  The price is either a reduced field of view, requiring a scanning strategy until the SBO is identified, or a significantly more expensive detection module.  In theory, existing wide-field IR instruments can make such surveys in daytime if narrow band filters can sufficiently reduce the count rates and there is a means of maintaining thermal stability, but the control, scheduling, and safety concerns for these facilities are likely problematic.

Moreover, a SBO detection system run by a neutrino detector group can simply be regarded as part of the overall system, allowing it to internally operate at a triggering rate viewed as unacceptably high for public announcements.

\section{Observing Extragalactic Shock Breakouts}
\label{app:Two}
Megaton-class neutrino detectors are capable of detecting
small numbers of neutrinos from ccSN out to roughly 10~Mpc \citep{ando05,kistler11}.  The ccSN rate at these distances is dominated by roughly 40
galaxies, representing 90\% of the local rate of 1-2/year \citep{ando05}.  The
key issue for flagging a small number of neutrino events is to narrow the 
temporal search window so that there is a negligible probability 
of background events in the window. Narrowing the temporal search window would similarly improve the sensitivity of gravitational wave detectors.  \cite{cowen10} show that given a well-sampled early-time
light curve, particularly one which has some information on the
shock breakout signal, the time of the core collapse can be estimated to within hours.  The problem, however, is that the short SBO
durations mean that normal surveys are unlikely to catch the breakout pulse except by chance.  In this sense, the \cite{kistler12}
emphasis on the importance of finding shock breakouts has the problem
backwards -- you want to use a high sensitivity survey for breakout pulses to trigger searches in the low sensitivity neutrino or
gravity wave detectors rather than the reverse.

It is feasible to simply monitor all or some of these
galaxies for shock breakout events, but it requires a
fairly industrial approach to the problem.  At 10~Mpc, and again 
assuming the radiation thermalizes, the $\sim16$~mag optical SBO events 
(Fig. \ref{fig:sbo_mag}) are easily accessible using off-the-shelf
equipment.  For example, a 12.5in Planewave telescope   
with an SBIG STXL-1102 camera has a roughly $0.5\times1.0$ degree
field of view that basically covers all the relevant 
galaxies except M~31 and M~33 in a single exposure and
would have $S/N\sim 5$ at $R\sim 20$ in 60 seconds for
a price on the order \$35,000 per unit after including
a Paramount ME robotic mount and a control computer. 
Based on the Winer Observatory, an additional \$10,000/year
would provide space in an existing dome and basic servicing
needs. Thus, for an overall system with $32$ units, the
hardware costs are roughly \$1,000,000 (assuming a modest
discount from list prices given the scale of the order)
with direct operating costs of order \$300,000/year.

The telescopes need to be sited relatively uniformly in 
longitude, with a greater emphasis on the North.  While
we did not attempt a detailed optimization, we experimented
with the achievable completeness defined by the fraction of
the local SN rate which could be monitored at a given cadence.
The rates for each galaxy were set following \cite{cappellaro99}
 based on the galaxy type and absolute magnitude.  We
located equal numbers of telescopes at 8 existing observatory sites
(Canaries, Cerro Tololo, Hawaii, Kitt Peak, Maidanak, Siding Spring, 
Sutherland and Xinglong station) and assumed a random and
uncorrelated 30\% of days were cloudy. Telescopes were assigned
to the un-observed, visible (after $18^\circ$ twilight, airmass $<2$) galaxy
with the highest estimated SN rate.  For a one minute
cadence, needed to try to catch breakout events from blue
supergiants (SN~87A, Type~IIb, Type~Ib), systems with 8,
16, 24, 32 and 40 telescopes achieved 11, 20, 26, 31 and 35\%
completeness.  For a 5 minute cadence, which would miss most
breakouts from blue supergiants but monitor those from red
supergiants well, the fractions are 19, 33, 40, 45 and 48\%.
In practice, systems with larger numbers of telescopes have
significant ``idle time'' on a strict 5 minute cadence, so
there would be significant numbers of observations on shorter
time baselines.  With 8 sites, our 30\% uncorrelated weather 
losses have negligible effects on the coverage fraction.

It is clear, however, that an optimized design would use still
more sites since allowing observations all the way down to
the horizon raises the (5 minute) fractions to 31, 51, 61, 69
and 74\% respectively.  Since there is no need for superb
image quality (darkness, cloud cover and bandwidth are 
the more relevant criteria), we will assume that an optimized
system of 32 telescopes can achieve 60\% coverage for Type~II 
(red supergiant) breakout shocks.  While some breakout shocks 
from blue or Wolf-Rayet stars will be detected, they would not 
be well sampled.  Since Type~II SNe then represent $\sim 60\%$
of the overall SN rate \citep{li11a}, the system would 
detect shock breakouts from roughly $1/3$ of local SN. The
average rate from these galaxies is somewhat uncertain, but is 
in the range of $1$-$2$~SN/year, so the system would detect
one breakout event every $\sim 2$-$3$ years.  While the yield is far lower than the UV spacecraft proposed by \cite{sagiv13}, the costs are also much lower and the system will find the closest SN, for which there is the greatest likelihood of a neutrino detection.
\end{appendix}

\bibliographystyle{apj}

\begin{thebibliography}{132}
\expandafter\ifx\csname natexlab\endcsname\relax\def\natexlab#1{#1}\fi

\bibitem[{{Abbasi} {et~al.}(2011){Abbasi}, {Abdou}, {Abu-Zayyad}, {Ackermann},
  {Adams}, {Aguilar}, {Ahlers}, {Allen}, {Altmann}, {Andeen}, \&
  et~al.}]{abbasi11}
{Abbasi}, R., {et~al.} 2011, \aap, 535, A109

\bibitem[{{Ageorges} {et~al.}(2010){Ageorges}, {Seifert}, {J{\"u}tte},
  {Knierim}, {Lehmitz}, {Germeroth}, {Buschkamp}, {Polsterer}, {Pasquali},
  {Naranjo}, {Gemperlein}, {Hill}, {Feiz}, {Hofmann}, {Laun}, {Lederer},
  {Lenzen}, {Mall}, {Mandel}, {M{\"u}ller}, {Quirrenbach}, {Sch{\"a}ffner},
  {Storz}, \& {Weiser}}]{ageorges10}
{Ageorges}, N., {et~al.} 2010, Proc. SPIE, 7735, 77351L

\bibitem[{{Alekseev} {et~al.}(1987){Alekseev}, {Alekseeva}, {Volchenko}, \&
  {Krivosheina}}]{alekseev87}
{Alekseev}, E.~N., {Alekseeva}, L.~N., {Volchenko}, V.~I., \& {Krivosheina},
  I.~V. 1987, PZETF, 45, 461

\bibitem[{{Alexeyev} \& {Alexeyeva}(2002)}]{alexeyev02}
{Alexeyev}, E.~N., \& {Alexeyeva}, L.~N. 2002, JETP, 95, 5

\bibitem[{{Andersson} {et~al.}(2013){Andersson}, {Baker}, {Belczynski},
  {Bernuzzi}, {Berti}, {Cadonati}, {Cerda-Duran}, {Clark}, {Favata}, {Finn},
  {Fryer}, {Giacomazzo}, {Gonzalez}, {Hendry}, {Heng}, {Hild},
  {Johnson-McDaniel}, {Kalmus}, {Klimenko}, {Kobayashi}, {Kokkotas}, {Laguna},
  {Lehner}, {Levin}, {Liebling}, {MacFadyen}, {Mandel}, {Marka}, {Marka},
  {Neilsen}, {O'Brien}, {Perna}, {Pfeiffer}, {Read}, {Reisswig}, {Rodriguez},
  {Ruffert}, {Schnetter}, {Searle}, {Shawhan}, {Shoemaker}, {Soderberg},
  {Sperhake}, {Sutton}, {Tanvir}, {Was}, \& {Whitcomb}}]{andersson13}
{Andersson}, N., {et~al.} 2013, CQGra, 30, 193002

\bibitem[{{Ando} {et~al.}(2005){Ando}, {Beacom}, \& {Y{\"u}ksel}}]{ando05}
{Ando}, S., {Beacom}, J.~F., \& {Y{\"u}ksel}, H. 2005, PhRvL,
  95, 171101

\bibitem[{{Antonioli} {et~al.}(2004){Antonioli}, {Tresch Fienberg}, {Fleurot},
  {Fukuda}, {Fulgione}, {Habig}, {Heise}, {McDonald}, {Mills}, {Namba},
  {Robinson}, {Scholberg}, {Schwendener}, {Sinnott}, {Stacey}, {Suzuki},
  {Tafirout}, {Vigorito}, {Viren}, {Virtue}, \& {Zichichi}}]{antonioli04}
{Antonioli}, P., {et~al.} 2004, NJPh, 6, 114

\bibitem[{{Arce} \& {Goodman}(1999)}]{arce99}
{Arce}, H.~G., \& {Goodman}, A.~A. 1999, \apjl, 512, L135

\bibitem[{{Arnett} {et~al.}(1989){Arnett}, {Bahcall}, {Kirshner}, \&
  {Woosley}}]{arnett89}
{Arnett}, W.~D., {Bahcall}, J.~N., {Kirshner}, R.~P., \& {Woosley}, S.~E. 1989,
  \araa, 27, 629

\bibitem[{{Beacom} {et~al.}(2001){Beacom}, {Boyd}, \& {Mezzacappa}}]{beacom01}
{Beacom}, J.~F., {Boyd}, R.~N., \& {Mezzacappa}, A. 2001, \prd, 63, 073011

\bibitem[{{Beacom} \& {Vagins}(2004)}]{Beacom04}
{Beacom}, J.~F., \& {Vagins}, M.~R. 2004, PhRvL, 93, 171101

\bibitem[{{Beacom} \& {Vogel}(1999)}]{beacom99}
{Beacom}, J.~F., \& {Vogel}, P. 1999, \prd, 60, 033007

\bibitem[{{Bionta} {et~al.}(1987){Bionta}, {Blewitt}, {Bratton}, {Casper}, \&
  {Ciocio}}]{bionta87}
{Bionta}, R.~M., {Blewitt}, G., {Bratton}, C.~B., {Casper}, D., \& {Ciocio}, A.
  1987, PhRvL, 58, 1494

\bibitem[{{Bonifacio} {et~al.}(2000){Bonifacio}, {Monai}, \&
  {Beers}}]{bonifacio00}
{Bonifacio}, P., {Monai}, S., \& {Beers}, T.~C. 2000, \aj, 120, 2065

\bibitem[{{Brandt} {et~al.}(2010){Brandt}, {Tojeiro}, {Aubourg}, {Heavens},
  {Jimenez}, \& {Strauss}}]{brandt10}
{Brandt}, T.~D., {Tojeiro}, R., {Aubourg}, {\'E}., {Heavens}, A., {Jimenez},
  R., \& {Strauss}, M.~A. 2010, \aj, 140, 804

\bibitem[{{Burrows}(2013)}]{burrows13}
{Burrows}, A. 2013, Reviews of Modern Physics, 85, 245

\bibitem[{{Campana} {et~al.}(2006){Campana}, {Mangano}, {Blustin}, {Brown},
  {Burrows}, {Chincarini}, {Cummings}, {Cusumano}, {Della Valle}, {Malesani},
  {M{\'e}sz{\'a}ros}, {Nousek}, {Page}, {Sakamoto}, {Waxman}, {Zhang}, {Dai},
  {Gehrels}, {Immler}, {Marshall}, {Mason}, {Moretti}, {O'Brien}, {Osborne},
  {Page}, {Romano}, {Roming}, {Tagliaferri}, {Cominsky}, {Giommi}, {Godet},
  {Kennea}, {Krimm}, {Angelini}, {Barthelmy}, {Boyd}, {Palmer}, {Wells}, \&
  {White}}]{campana06}
{Campana}, S., {et~al.} 2006, \nat, 442, 1008

\bibitem[{{Cappellaro} {et~al.}(1999){Cappellaro}, {Evans}, \&
  {Turatto}}]{cappellaro99}
{Cappellaro}, E., {Evans}, R., \& {Turatto}, M. 1999, \aap, 351, 459

\bibitem[{{Cardelli} {et~al.}(1989){Cardelli}, {Clayton}, \&
  {Mathis}}]{cardelli89}
{Cardelli}, J.~A., {Clayton}, G.~C., \& {Mathis}, J.~S. 1989, \apj, 345, 245

\bibitem[{{Chen} {et~al.}(1999){Chen}, {Figueras}, {Torra}, {Jordi}, {Luri}, \&
  {Galad{\'{\i}}-Enr{\'{\i}}quez}}]{chen99}
{Chen}, B., {Figueras}, F., {Torra}, J., {Jordi}, C., {Luri}, X., \&
  {Galad{\'{\i}}-Enr{\'{\i}}quez}, D. 1999, \aap, 352, 459

\bibitem[{{Chomiuk} \& {Povich}(2011)}]{chomiuk11}
{Chomiuk}, L., \& {Povich}, M.~S. 2011, \aj, 142, 197

\bibitem[{{Clark} \& {Stephenson}(1977)}]{clark77}
{Clark}, D.~H., \& {Stephenson}, F.~R. 1977, {The Historical Supernovae} (Oxford: Pergamon)

\bibitem[{{Cowen} {et~al.}(2010){Cowen}, {Franckowiak}, \&
  {Kowalski}}]{cowen10}
{Cowen}, D.~F., {Franckowiak}, A., \& {Kowalski}, M. 2010, Astroparticle
  Physics, 33, 19

\bibitem[{{Davies} {et~al.}(2011){Davies}, {Hoare}, {Lumsden}, {Hosokawa},
  {Oudmaijer}, {Urquhart}, {Mottram}, \& {Stead}}]{davies11}
{Davies}, B., {Hoare}, M.~G., {Lumsden}, S.~L., {Hosokawa}, T., {Oudmaijer},
  R.~D., {Urquhart}, J.~S., {Mottram}, J.~C., \& {Stead}, J. 2011, \mnras, 416,
  972

\bibitem[{{Diehl}(2012)}]{diehl12}
{Diehl}, R. 2012, arXiv:1202.0481

\bibitem[{{Diehl} {et~al.}(2006){Diehl}, {Halloin}, {Kretschmer}, {Lichti},
  {Sch{\"o}nfelder}, {Strong}, {von Kienlin}, {Wang}, {Jean}, {Kn{\"o}dlseder},
  {Roques}, {Weidenspointner}, {Schanne}, {Hartmann}, {Winkler}, \&
  {Wunderer}}]{diehl06}
{Diehl}, R., {et~al.} 2006, \nat, 439, 45

\bibitem[{{Drake} {et~al.}(2009){Drake}, {Djorgovski}, {Mahabal}, {Beshore},
  {Larson}, {Graham}, {Williams}, {Christensen}, {Catelan}, {Boattini},
  {Gibbs}, {Hill}, \& {Kowalski}}]{drake09}
{Drake}, A.~J., {et~al.} 2009, \apj, 696, 870

\bibitem[{{Drew} {et~al.}(2005){Drew}, {Greimel}, {Irwin}, {Aungwerojwit},
  {Barlow}, {Corradi}, {Drake}, {G{\"a}nsicke}, {Groot}, {Hales}, {Hopewell},
  {Irwin}, {Knigge}, {Leisy}, {Lennon}, {Mampaso}, {Masheder}, {Matsuura},
  {Morales-Rueda}, {Morris}, {Parker}, {Phillipps}, {Rodriguez-Gil}, {Roelofs},
  {Skillen}, {Sokoloski}, {Steeghs}, {Unruh}, {Viironen}, {Vink}, {Walton},
  {Witham}, {Wright}, {Zijlstra}, \& {Zurita}}]{drew05}
{Drew}, J.~E., {et~al.} 2005, \mnras, 362, 753

\bibitem[{{Edwards} {et~al.}(2012){Edwards}, {Pagnotta}, \&
  {Schaefer}}]{edwards12}
{Edwards}, Z.~I., {Pagnotta}, A., \& {Schaefer}, B.~E. 2012, \apjl, 747, L19

\bibitem[{{Eldridge} {et~al.}(2013){Eldridge}, {Fraser}, {Smartt}, {Maund}, \&
  {Crockett}}]{eldridge13}
{Eldridge}, J.~J., {Fraser}, M., {Smartt}, S.~J., {Maund}, J.~R., \&
  {Crockett}, R.~M. 2013, \mnras, 2346

\bibitem[{{Faucher-Gigu{\`e}re} \& {Kaspi}(2006)}]{faucher06}
{Faucher-Gigu{\`e}re}, C.-A., \& {Kaspi}, V.~M. 2006, \apj, 643, 332

\bibitem[{{Folatelli} {et~al.}(2010){Folatelli}, {Phillips}, {Burns},
  {Contreras}, {Hamuy}, {Freedman}, {Persson}, {Stritzinger}, {Suntzeff},
  {Krisciunas}, {Boldt}, {Gonz{\'a}lez}, {Krzeminski}, {Morrell}, {Roth},
  {Salgado}, {Madore}, {Murphy}, {Wyatt}, {Li}, {Filippenko}, \&
  {Miller}}]{folatelli10}
{Folatelli}, G., {et~al.} 2010, \aj, 139, 120

\bibitem[{{Fryxell} {et~al.}(1991){Fryxell}, {Arnett}, \&
  {Mueller}}]{fryxell91}
{Fryxell}, B., {Arnett}, D., \& {Mueller}, E. 1991, \apj, 367, 619

\bibitem[{Fukuda {et~al.}(2003)}]{fukuda03}
Fukuda, Y., {et~al.} 2003, Nucl.Instrum.Meth., A501, 418

\bibitem[{{Gal-Yam} \& {Leonard}(2009)}]{galyam09}
{Gal-Yam}, A., \& {Leonard}, D.~C. 2009, \nat, 458, 865

\bibitem[{{Gehrels} {et~al.}(1987){Gehrels}, {Leventhal}, \&
  {MacCallum}}]{gehrels87}
{Gehrels}, N., {Leventhal}, M., \& {MacCallum}, C.~J. 1987, \apj, 322, 215

\bibitem[{{Girardi} {et~al.}(2005){Girardi}, {Groenewegen}, {Hatziminaoglou},
  \& {da Costa}}]{girardi05}
{Girardi}, L., {Groenewegen}, M.~A.~T., {Hatziminaoglou}, E., \& {da Costa}, L.
  2005, \aap, 436, 895

\bibitem[{{Gould}(2013)}]{gould13}
{Gould}, A. 2013, arXiv:1304.3455

\bibitem[{{Groh} {et~al.}(2013){Groh}, {Meynet}, {Georgy}, \&
  {Ekstrom}}]{groh13}
{Groh}, J.~H., {Meynet}, G., {Georgy}, C., \& {Ekstrom}, S. 2013,
  \aap, 558, A131

\bibitem[{{Hamuy}(2003)}]{hamuy03}
{Hamuy}, M. 2003, \apj, 582, 905

\bibitem[{{Heger} {et~al.}(2003){Heger}, {Fryer}, {Woosley}, {Langer}, \&
  {Hartmann}}]{heger03}
{Heger}, A., {Fryer}, C.~L., {Woosley}, S.~E., {Langer}, N., \& {Hartmann},
  D.~H. 2003, \apj, 591, 288

\bibitem[{{Hirata} {et~al.}(1987){Hirata}, {Kajita}, {Koshiba}, {Nakahata}, \&
  {Oyama}}]{hirata87}
{Hirata}, K., {Kajita}, T., {Koshiba}, M., {Nakahata}, M., \& {Oyama}, Y. 1987,
  PhRvL, 58, 1490

\bibitem[{{Horiuchi} \& {Beacom}(2010)}]{horiuchi10}
{Horiuchi}, S., \& {Beacom}, J.~F. 2010, \apj, 723, 329

\bibitem[{{Horiuchi} {et~al.}(2011){Horiuchi}, {Beacom}, {Kochanek}, {Prieto},
  {Stanek}, \& {Thompson}}]{horiuchi11}
{Horiuchi}, S., {Beacom}, J.~F., {Kochanek}, C.~S., {Prieto}, J.~L., {Stanek},
  K.~Z., \& {Thompson}, T.~A. 2011, \apj, 738, 154

\bibitem[{{Hungerford} {et~al.}(2003){Hungerford}, {Fryer}, \&
  {Warren}}]{hungerford03}
{Hungerford}, A.~L., {Fryer}, C.~L., \& {Warren}, M.~S. 2003, \apj, 594, 390

\bibitem[{{Ikeda} {et~al.}(2007){Ikeda}, {Takeda}, {Fukuda}, {Vagins}, {Abe},
  {Iida}, {Ishihara}, {Kameda}, {Koshio}, {Minamino}, {Mitsuda}, {Miura},
  {Moriyama}, {Nakahata}, {Obayashi}, {Ogawa}, {Sekiya}, {Shiozawa}, {Suzuki},
  {Takeuchi}, {Ueshima}, {Watanabe}, {Yamada}, {Higuchi}, {Ishihara},
  {Ishitsuka}, {Kajita}, {Kaneyuki}, {Mitsuka}, {Nakayama}, {Nishino},
  {Okumura}, {Saji}, {Takenaga}, {Clark}, {Desai}, {Dufour}, {Kearns},
  {Likhoded}, {Litos}, {Raaf}, {Stone}, {Sulak}, {Wang}, {Goldhaber}, {Casper},
  {Cravens}, {Dunmore}, {Kropp}, {Liu}, {Mine}, {Regis}, {Smy}, {Sobel},
  {Ganezer}, {Hill}, {Keig}, {Jang}, {Kim}, {Lim}, {Scholberg}, {Tanimoto},
  {Walter}, {Wendell}, {Ellsworth}, {Tasaka}, {Guillian}, {Learned}, {Matsuno},
  {Messier}, {Hayato}, {Ichikawa}, {Ishida}, {Ishii}, {Iwashita}, {Kobayashi},
  {Nakadaira}, {Nakamura}, {Nitta}, {Oyama}, {Totsuka}, {Suzuki}, {Hasegawa},
  {Hiraide}, {Maesaka}, {Nakaya}, {Nishikawa}, {Sasaki}, {Yamamoto},
  {Yokoyama}, {Haines}, {Dazeley}, {Hatakeyama}, {Svoboda}, {Sullivan},
  {Turcan}, {Habig}, {Sato}, {Itow}, {Koike}, {Tanaka}, {Jung}, {Kato},
  {Kobayashi}, {Malek}, {McGrew}, {Sarrat}, {Terri}, {Yanagisawa}, {Tamura},
  {Idehara}, {Sakuda}, {Sugihara}, {Kuno}, {Yoshida}, {Kim}, {Yang}, {Yoo},
  {Ishizuka}, {Okazawa}, {Choi}, {Seo}, {Gando}, {Hasegawa}, {Inoue}, {Furuse},
  {Ishii}, {Nishijima}, {Ishino}, {Watanabe}, {Koshiba}, {Chen}, {Deng}, {Liu},
  {Kielczewska}, {Zalipska}, {Berns}, {Gran}, {Shiraishi}, {Stachyra},
  {Thrane}, {Washburn}, {Wilkes}, \& {Super-KAMIOKANDE
  Collaboration}}]{ikeda07}
{Ikeda}, M., {et~al.} 2007, \apj, 669, 519

\bibitem[{{Janka}(2012)}]{janka12}
{Janka}, H.-T. 2012, ARNPS, 62, 407

\bibitem[{{Jenniskens} \& {Greenberg}(1993)}]{jenniskens93}
{Jenniskens}, P., \& {Greenberg}, J.~M. 1993, \aap, 274, 439

\bibitem[{{Jim} {et~al.}(2011){Jim}, {Gibson}, \& {Pier}}]{jim11}
{Jim}, K., {Gibson}, B., \& {Pier}, E. 2011, in Advanced Maui Optical and Space
  Surveillance Technologies Conference

\bibitem[{{Keane} \& {Kramer}(2008)}]{keane08}
{Keane}, E.~F., \& {Kramer}, M. 2008, \mnras, 391, 2009

\bibitem[{{Kerzendorf} {et~al.}(2012){Kerzendorf}, {Schmidt}, {Laird},
  {Podsiadlowski}, \& {Bessell}}]{kerzendorf12}
{Kerzendorf}, W.~E., {Schmidt}, B.~P., {Laird}, J.~B., {Podsiadlowski}, P., \&
  {Bessell}, M.~S. 2012, \apj, 759, 7

\bibitem[{{Kistler} {et~al.}(2012){Kistler}, {Haxton}, \& {Yuksel}}]{kistler12}
{Kistler}, M.~D., {Haxton}, W.~C., \& {Yuksel}, H. 2012, arXiv:1211.6770

\bibitem[{{Kistler} {et~al.}(2011){Kistler}, {Y{\"u}ksel}, {Ando}, {Beacom}, \&
  {Suzuki}}]{kistler11}
{Kistler}, M.~D., {Y{\"u}ksel}, H., {Ando}, S., {Beacom}, J.~F., \& {Suzuki},
  Y. 2011, \prd, 83, 123008

\bibitem[{{Kochanek} {et~al.}(2008){Kochanek}, {Beacom}, {Kistler}, {Prieto},
  {Stanek}, {Thompson}, \& {Y{\"u}ksel}}]{kochanek08}
{Kochanek}, C.~S., {Beacom}, J.~F., {Kistler}, M.~D., {Prieto}, J.~L.,
  {Stanek}, K.~Z., {Thompson}, T.~A., \& {Y{\"u}ksel}, H. 2008, \apj, 684, 1336

\bibitem[{{Krisciunas} {et~al.}(2009){Krisciunas}, {Hamuy}, {Suntzeff},
  {Espinoza}, {Gonzalez}, {Gonzalez}, {Gonzalez}, {Koviak}, {Krzeminski},
  {Morrell}, {Phillips}, {Roth}, \& {Thomas-Osip}}]{krisciunas09}
{Krisciunas}, K., {et~al.} 2009, \aj, 137, 34

\bibitem[{{Kunkel} {et~al.}(1987){Kunkel}, {Madore}, {Shelton}, {Duhalde},
  {Bateson}, {Jones}, {Moreno}, {Walker}, {Garradd}, {Warner}, \&
  {Menzies}}]{kunkel87}
{Kunkel}, W., {et~al.} 1987, \iaucirc, 4316, 1

\bibitem[{{Law} {et~al.}(2009){Law}, {Kulkarni}, {Dekany}, {Ofek}, {Quimby},
  {Nugent}, {Surace}, {Grillmair}, {Bloom}, {Kasliwal}, {Bildsten}, {Brown},
  {Cenko}, {Ciardi}, {Croner}, {Djorgovski}, {van Eyken}, {Filippenko}, {Fox},
  {Gal-Yam}, {Hale}, {Hamam}, {Helou}, {Henning}, {Howell}, {Jacobsen},
  {Laher}, {Mattingly}, {McKenna}, {Pickles}, {Poznanski}, {Rahmer}, {Rau},
  {Rosing}, {Shara}, {Smith}, {Starr}, {Sullivan}, {Velur}, {Walters}, \&
  {Zolkower}}]{law09}
{Law}, N.~M., {et~al.} 2009, \pasp, 121, 1395

\bibitem[{{Leaman} {et~al.}(2011){Leaman}, {Li}, {Chornock}, \&
  {Filippenko}}]{leaman11}
{Leaman}, J., {Li}, W., {Chornock}, R., \& {Filippenko}, A.~V. 2011, \mnras,
  412, 1419

\bibitem[{{Leonor} {et~al.}(2010){Leonor}, {Cadonati}, {Coccia}, {D'Antonio},
  {Di Credico}, {Fafone}, {Frey}, {Fulgione}, {Katsavounidis}, {Ott},
  {Pagliaroli}, {Scholberg}, {Thrane}, \& {Vissani}}]{leonor10}
{Leonor}, I., {et~al.} 2010, CQGra, 27, 084019

\bibitem[{{Li} {et~al.}(2011{\natexlab{a}}){Li}, {Chornock}, {Leaman},
  {Filippenko}, {Poznanski}, {Wang}, {Ganeshalingam}, \& {Mannucci}}]{li11b}
{Li}, W., {Chornock}, R., {Leaman}, J., {Filippenko}, A.~V., {Poznanski}, D.,
  {Wang}, X., {Ganeshalingam}, M., \& {Mannucci}, F. 2011{\natexlab{a}},
  \mnras, 412, 1473

\bibitem[{{Li} {et~al.}(2011{\natexlab{b}}){Li}, {Leaman}, {Chornock},
  {Filippenko}, {Poznanski}, {Ganeshalingam}, {Wang}, {Modjaz}, {Jha}, {Foley},
  \& {Smith}}]{li11a}
{Li}, W., {et~al.} 2011{\natexlab{b}}, \mnras, 412, 1441

\bibitem[{{Lien} {et~al.}(2010){Lien}, {Fields}, \& {Beacom}}]{lien10}
{Lien}, A., {Fields}, B.~D., \& {Beacom}, J.~F. 2010, \prd, 81, 083001

\bibitem[{{Lovegrove} \& {Woosley}(2013)}]{lovegrove13}
{Lovegrove}, E., \& {Woosley}, S.~E. 2013, \apj, 769, 109

\bibitem[{{LSST Dark Energy Science Collaboration}(2012)}]{abate12}
{LSST Dark Energy Science Collaboration}. 2012, arXiv:1211.0310

\bibitem[{{Majewski} {et~al.}(2011){Majewski}, {Zasowski}, \&
  {Nidever}}]{majewski11}
{Majewski}, S.~R., {Zasowski}, G., \& {Nidever}, D.~L. 2011, \apj, 739, 25

\bibitem[{{Mannucci} {et~al.}(2006){Mannucci}, {Della Valle}, \&
  {Panagia}}]{mannucci06}
{Mannucci}, F., {Della Valle}, M., \& {Panagia}, N. 2006, \mnras, 370, 773

\bibitem[{{Maoz} {et~al.}(2012){Maoz}, {Mannucci}, \& {Brandt}}]{maoz12}
{Maoz}, D., {Mannucci}, F., \& {Brandt}, T.~D. 2012, \mnras, 426, 3282

\bibitem[{{Marek} \& {Janka}(2009)}]{marek09}
{Marek}, A., \& {Janka}, H.-T. 2009, \apj, 694, 664

\bibitem[{{Marigo} {et~al.}(2008){Marigo}, {Girardi}, {Bressan}, {Groenewegen},
  {Silva}, \& {Granato}}]{marigo08}
{Marigo}, P., {Girardi}, L., {Bressan}, A., {Groenewegen}, M.~A.~T., {Silva},
  L., \& {Granato}, G.~L. 2008, \aap, 482, 883

\bibitem[{{Matz} {et~al.}(1988){Matz}, {Share}, {Leising}, {Chupp}, \&
  {Vestrand}}]{matz88}
{Matz}, S.~M., {Share}, G.~H., {Leising}, M.~D., {Chupp}, E.~L., \& {Vestrand},
  W.~T. 1988, \nat, 331, 416

\bibitem[{{Matzner} \& {McKee}(1999)}]{matzner99}
{Matzner}, C.~D., \& {McKee}, C.~F. 1999, \apj, 510, 379

\bibitem[{{Mauerhan} {et~al.}(2013){Mauerhan}, {Smith}, {Filippenko},
  {Blanchard}, {Blanchard}, {Casper}, {Cenko}, {Clubb}, {Cohen}, {Fuller},
  {Li}, \& {Silverman}}]{mauerhan13}
{Mauerhan}, J.~C., {et~al.} 2013, \mnras, 430, 1801

\bibitem[{{Maund} {et~al.}(2013){Maund}, {Reilly}, \& {Mattila}}]{maund13}
{Maund}, J., {Reilly}, E., \& {Mattila}, S. 2013, arXiv:1302.7152

\bibitem[{{McCray}(1993)}]{mccray93}
{McCray}, R. 1993, \araa, 31, 175

\bibitem[{{McKee} \& {Williams}(1997)}]{mckee97}
{McKee}, C.~F., \& {Williams}, J.~P. 1997, \apj, 476, 144

\bibitem[{{Mezzacappa}(2005)}]{mezzacappa05}
{Mezzacappa}, A. 2005, ARNPS, 55, 467

\bibitem[{{Minniti} {et~al.}(2010){Minniti}, {Lucas}, {Emerson}, {Saito},
  {Hempel}, {Pietrukowicz}, {Ahumada}, {Alonso}, {Alonso-Garcia}, {Arias},
  {Bandyopadhyay}, {Barb{\'a}}, {Barbuy}, {Bedin}, {Bica}, {Borissova},
  {Bronfman}, {Carraro}, {Catelan}, {Clari{\'a}}, {Cross}, {de Grijs},
  {D{\'e}k{\'a}ny}, {Drew}, {Fari{\~n}a}, {Feinstein}, {Fern{\'a}ndez
  Laj{\'u}s}, {Gamen}, {Geisler}, {Gieren}, {Goldman}, {Gonzalez}, {Gunthardt},
  {Gurovich}, {Hambly}, {Irwin}, {Ivanov}, {Jord{\'a}n}, {Kerins}, {Kinemuchi},
  {Kurtev}, {L{\'o}pez-Corredoira}, {Maccarone}, {Masetti}, {Merlo},
  {Messineo}, {Mirabel}, {Monaco}, {Morelli}, {Padilla}, {Palma}, {Parisi},
  {Pignata}, {Rejkuba}, {Roman-Lopes}, {Sale}, {Schreiber}, {Schr{\"o}der},
  {Smith}, {Sodr{\'e}}, {Soto}, {Tamura}, {Tappert}, {Thompson}, {Toledo},
  {Zoccali}, \& {Pietrzynski}}]{minniti10}
{Minniti}, D., {et~al.} 2010, NewA, 15, 433

\bibitem[{{Monet} {et~al.}(2003){Monet}, {Levine}, {Canzian}, {Ables}, {Bird},
  {Dahn}, {Guetter}, {Harris}, {Henden}, {Leggett}, {Levison}, {Luginbuhl},
  {Martini}, {Monet}, {Munn}, {Pier}, {Rhodes}, {Riepe}, {Sell}, {Stone},
  {Vrba}, {Walker}, {Westerhout}, {Brucato}, {Reid}, {Schoening}, {Hartley},
  {Read}, \& {Tritton}}]{monet03}
{Monet}, D.~G., {et~al.} 2003, \aj, 125, 984

\bibitem[{{Murray} \& {Rahman}(2010)}]{murray10}
{Murray}, N., \& {Rahman}, M. 2010, \apj, 709, 424

\bibitem[{{Nadezhin}(1980)}]{nadezhin80}
{Nadezhin}, D.~K. 1980, \apss, 69, 115

\bibitem[{{Nakar} \& {Sari}(2010)}]{nakar10}
{Nakar}, E., \& {Sari}, R. 2010, \apj, 725, 904

\bibitem[{{Nakazato} {et~al.}(2012){Nakazato}, {Furusawa}, {Sumiyoshi},
  {Ohnishi}, {Yamada}, \& {Suzuki}}]{nakazato12}
{Nakazato}, K., {Furusawa}, S., {Sumiyoshi}, K., {Ohnishi}, A., {Yamada}, S.,
  \& {Suzuki}, H. 2012, \apj, 745, 197

\bibitem[{{Nakazato} {et~al.}(2007){Nakazato}, {Sumiyoshi}, \&
  {Yamada}}]{nakazato07}
{Nakazato}, K., {Sumiyoshi}, K., \& {Yamada}, S. 2007, \apj, 666, 1140

\bibitem[{{Nataf} {et~al.}(2013){Nataf}, {Gould}, {Fouqu{\'e}}, {Gonzalez},
  {Johnson}, {Skowron}, {Udalski}, {Szyma{\'n}ski}, {Kubiak},
  {Pietrzy{\'n}ski}, {Soszy{\'n}ski}, {Ulaczyk}, {Wyrzykowski}, \&
  {Poleski}}]{nataf13}
{Nataf}, D.~M., {et~al.} 2013, \apj, 769, 88

\bibitem[{{Ng} {et~al.}(2012){Ng}, {Horiuchi}, {Beacom}, {Siegal-Gaskins},
  {Preece}, {Smith}, \& {Gelbord}}]{ng12}
{Ng}, K.~C.~Y., {Horiuchi}, S., {Beacom}, J.~F., {Siegal-Gaskins}, J.,
  {Preece}, R., {Smith}, M., \& {Gelbord}, J. 2012, ATel,
  4473, 1

\bibitem[{{Nidever} {et~al.}(2012){Nidever}, {Zasowski}, \&
  {Majewski}}]{nidever12}
{Nidever}, D.~L., {Zasowski}, G., \& {Majewski}, S.~R. 2012, \apjs, 201, 35

\bibitem[{{O'Connor} \& {Ott}(2011)}]{oconnor11}
{O'Connor}, E., \& {Ott}, C.~D. 2011, \apj, 730, 70

\bibitem[{{Odrzywolek} \& {Plewa}(2011)}]{odrzywolek11}
{Odrzywolek}, A., \& {Plewa}, T. 2011, \aap, 529, A156

\bibitem[{{Ofek} {et~al.}(2013){Ofek}, {Sullivan}, {Cenko}, {Kasliwal},
  {Gal-Yam}, {Kulkarni}, {Arcavi}, {Bildsten}, {Bloom}, {Horesh}, {Howell},
  {Filippenko}, {Laher}, {Murray}, {Nakar}, {Nugent}, {Silverman}, {Shaviv},
  {Surace}, \& {Yaron}}]{ofek13}
{Ofek}, E.~O., {et~al.} 2013, \nat, 494, 65

\bibitem[{{Olofsson} \& {Olofsson}(2010)}]{olofsson10}
{Olofsson}, S., \& {Olofsson}, G. 2010, \aap, 522, A84

\bibitem[{{Ott}(2009)}]{ott09}
{Ott}, C.~D. 2009, CQGra, 26, 063001

\bibitem[{{Pastorello} {et~al.}(2007){Pastorello}, {Smartt}, {Mattila},
  {Eldridge}, {Young}, {Itagaki}, {Yamaoka}, {Navasardyan}, {Valenti}, {Patat},
  {Agnoletto}, {Augusteijn}, {Benetti}, {Cappellaro}, {Boles}, {Bonnet-Bidaud},
  {Botticella}, {Bufano}, {Cao}, {Deng}, {Dennefeld}, {Elias-Rosa},
  {Harutyunyan}, {Keenan}, {Iijima}, {Lorenzi}, {Mazzali}, {Meng}, {Nakano},
  {Nielsen}, {Smoker}, {Stanishev}, {Turatto}, {Xu}, \&
  {Zampieri}}]{pastorello07}
{Pastorello}, A., {et~al.} 2007, \nat, 447, 829

\bibitem[{{Pejcha} \& {Thompson}(2012)}]{pejcha12}
{Pejcha}, O., \& {Thompson}, T.~A. 2012, \apj, 746, 106

\bibitem[{{Pietrzy{\'n}ski} {et~al.}(2013){Pietrzy{\'n}ski}, {Graczyk},
  {Gieren}, {Thompson}, {Pilecki}, {Udalski}, {Soszy{\'n}ski}, {Koz{\l}owski},
  {Konorski}, {Suchomska}, {Bono}, {Moroni}, {Villanova}, {Nardetto},
  {Bresolin}, {Kudritzki}, {Storm}, {Gallenne}, {Smolec}, {Minniti}, {Kubiak},
  {Szyma{\'n}ski}, {Poleski}, {Wyrzykowski}, {Ulaczyk}, {Pietrukowicz},
  {G{\'o}rski}, \& {Karczmarek}}]{pietrzynski13}
{Pietrzy{\'n}ski}, G., {et~al.} 2013, \nat, 495, 76

\bibitem[{{Piro}(2013)}]{piro13}
{Piro}, A.~L. 2013, \apjl, 768, L14

\bibitem[{{Piro} {et~al.}(2010){Piro}, {Chang}, \& {Weinberg}}]{piro10}
{Piro}, A.~L., {Chang}, P., \& {Weinberg}, N.~N. 2010, \apj, 708, 598

\bibitem[{{Plotkin} \& {Clayton}(2004)}]{plotkin04}
{Plotkin}, R.~M., \& {Clayton}, G.~C. 2004, JAVSO, 32, 89

\bibitem[{{Pojmanski}(2002)}]{pojmanski02}
{Pojmanski}, G. 2002, AcA, 52, 397

\bibitem[{{Reed}(2005)}]{reed05}
{Reed}, B.~C. 2005, \aj, 130, 1652

\bibitem[{{Robitaille} \& {Whitney}(2010)}]{robitaille10}
{Robitaille}, T.~P., \& {Whitney}, B.~A. 2010, \apjl, 710, L11

\bibitem[{{Rork} {et~al.}(1982){Rork}, {Lin}, \& {Yakutis}}]{rork82}
{Rork}, E.~W., {Lin}, S.~S., \& {Yakutis}, A.~J. 1982, STIN, 83, 10098

\bibitem[{{Sagiv} {et~al.}(2013){Sagiv}, {Gal-Yam}, {Ofek}, {Waxman},
  {Aharonson}, {Nakar}, {Maoz}, {Trakhtenbrot}, {Kulkarni}, {Phinney}, {Topaz},
  {Beichman}, {Murthy}, \& {Worden}}]{sagiv13}
{Sagiv}, I., {et~al.} 2013, arXiv:1303.6194

\bibitem[{{Sako} {et~al.}(2008){Sako}, {Bassett}, {Becker}, {Cinabro},
  {DeJongh}, {Depoy}, {Dilday}, {Doi}, {Frieman}, {Garnavich}, {Hogan},
  {Holtzman}, {Jha}, {Kessler}, {Konishi}, {Lampeitl}, {Marriner}, {Miknaitis},
  {Nichol}, {Prieto}, {Riess}, {Richmond}, {Romani}, {Schneider}, {Smith},
  {Subba Rao}, {Takanashi}, {Tokita}, {van der Heyden}, {Yasuda}, {Zheng},
  {Barentine}, {Brewington}, {Choi}, {Dembicky}, {Harnavek}, {Ihara}, {Im},
  {Ketzeback}, {Kleinman}, {Krzesi{\'n}ski}, {Long}, {Malanushenko},
  {Malanushenko}, {McMillan}, {Morokuma}, {Nitta}, {Pan}, {Saurage}, \&
  {Snedden}}]{sako08}
{Sako}, M., {et~al.} 2008, \aj, 135, 348

\bibitem[{{Sandie} {et~al.}(1988){Sandie}, {Nakano}, {Chase}, {Fishman},
  {Meegan}, {Wilson}, {Paciesas}, \& {Lasche}}]{sandie88}
{Sandie}, W.~G., {Nakano}, G.~H., {Chase}, Jr., L.~F., {Fishman}, G.~J.,
  {Meegan}, C.~A., {Wilson}, R.~B., {Paciesas}, W.~S., \& {Lasche}, G.~P. 1988,
  \apjl, 334, L91

\bibitem[{{Sapir} {et~al.}(2013){Sapir}, {Katz}, \& {Waxman}}]{sapir13}
{Sapir}, N., {Katz}, B., \& {Waxman}, E. 2013, \apj, 774, 79

\bibitem[{{Schaefer} \& {Pagnotta}(2012)}]{schaefer12}
{Schaefer}, B.~E., \& {Pagnotta}, A. 2012, \nat, 481, 164

\bibitem[{{Schawinski} {et~al.}(2008){Schawinski}, {Justham}, {Wolf},
  {Podsiadlowski}, {Sullivan}, {Steenbrugge}, {Bell}, {R{\"o}ser}, {Walker},
  {Astier}, {Balam}, {Balland}, {Carlberg}, {Conley}, {Fouchez}, {Guy},
  {Hardin}, {Hook}, {Howell}, {Pain}, {Perrett}, {Pritchet}, {Regnault}, \&
  {Yi}}]{schawinski08}
{Schawinski}, K., {et~al.} 2008, Science, 321, 223

\bibitem[{{Schlegel} {et~al.}(1998){Schlegel}, {Finkbeiner}, \&
  {Davis}}]{schlegel98}
{Schlegel}, D.~J., {Finkbeiner}, D.~P., \& {Davis}, M. 1998, \apj, 500, 525

\bibitem[{{Scholberg}(2012)}]{scholberg12}
{Scholberg}, K. 2012, ARNPS, 62, 81

\bibitem[{{Shappee} {et~al.}(2013){Shappee}, {Prieto}, {Grupe}, {Kochanek},
  {Stanek}, {De Rosa}, {Mathur}, {Zu}, {Peterson}, {Pogge}, {Komossa}, {Im},
  {Jencson}, {W-S.~Holoien}, {Basu}, {Beacom}, {Szczygiel}, {Brimacombe},
  {Adams}, {Campillay}, {Choi}, {Contreras}, {Dietrich}, {Dubberley},
  {Elphick}, {Foale}, {Giustini}, {Gonzalez}, {Hawkins}, {Howell}, {Hsiao},
  {Koss}, {Leighly}, {Morrell}, {Mudd}, {Mullins}, {Nugent}, {Parrent},
  {Phillips}, {Pojmanski}, {Rosing}, {Ross}, {Sand}, {Terndrup}, {Valenti},
  {Walker}, \& {Yoon}}]{shappee13}
{Shappee}, B.~J., {et~al.} 2013, arXiv:1310.2241

\bibitem[{{Skrutskie} {et~al.}(2006){Skrutskie}, {Cutri}, {Stiening},
  {Weinberg}, {Schneider}, {Carpenter}, {Beichman}, {Capps}, {Chester},
  {Elias}, {Huchra}, {Liebert}, {Lonsdale}, {Monet}, {Price}, {Seitzer},
  {Jarrett}, {Kirkpatrick}, {Gizis}, {Howard}, {Evans}, {Fowler}, {Fullmer},
  {Hurt}, {Light}, {Kopan}, {Marsh}, {McCallon}, {Tam}, {Van Dyk}, \&
  {Wheelock}}]{skrutskie06}
{Skrutskie}, M.~F., {et~al.} 2006, \aj, 131, 1163

\bibitem[{{Smartt}(2009)}]{smartt09r}
{Smartt}, S.~J. 2009, \araa, 47, 63

\bibitem[{{Smartt} {et~al.}(2009){Smartt}, {Eldridge}, {Crockett}, \&
  {Maund}}]{smartt09a}
{Smartt}, S.~J., {Eldridge}, J.~J., {Crockett}, R.~M., \& {Maund}, J.~R. 2009,
  \mnras, 395, 1409

\bibitem[{{Soderberg} {et~al.}(2008){Soderberg}, {Berger}, {Page}, {Schady},
  {Parrent}, {Pooley}, {Wang}, {Ofek}, {Cucchiara}, {Rau}, {Waxman}, {Simon},
  {Bock}, {Milne}, {Page}, {Barentine}, {Barthelmy}, {Beardmore}, {Bietenholz},
  {Brown}, {Burrows}, {Burrows}, {Byrngelson}, {Cenko}, {Chandra}, {Cummings},
  {Fox}, {Gal-Yam}, {Gehrels}, {Immler}, {Kasliwal}, {Kong}, {Krimm},
  {Kulkarni}, {Maccarone}, {M{\'e}sz{\'a}ros}, {Nakar}, {O'Brien}, {Overzier},
  {de Pasquale}, {Racusin}, {Rea}, \& {York}}]{soderberg08}
{Soderberg}, A.~M., {et~al.} 2008, \nat, 453, 469

\bibitem[{{Sokolovsky} {et~al.}(2013){Sokolovsky}, {Korotkiy}, \&
  {Lebedev}}]{sokolovsky13}
{Sokolovsky}, K., {Korotkiy}, S., \& {Lebedev}, A. 2013, arXiv:1303.3268

\bibitem[{{Stanek}(1998)}]{stanek98}
{Stanek}, K.~Z. 1998, arXiv:astro-ph/9802307

\bibitem[{{Stephenson} \& {Green}(2002)}]{stephenson02}
{Stephenson}, F.~R., \& {Green}, D.~A. 2002, Historical Supernovae and Their
  Remnants, International Series in Astronomy and Astrophysics, vol.~5 (Oxford: Clarendon)

\bibitem[{{Strom}(1994)}]{strom94}
{Strom}, R.~G. 1994, \aap, 288, L1

\bibitem[{{Suntzeff} \& {Bouchet}(1990)}]{suntzeff90}
{Suntzeff}, N.~B., \& {Bouchet}, P. 1990, \aj, 99, 650

\bibitem[{{Szczygie{\l}} {et~al.}(2012){Szczygie{\l}}, {Gerke}, {Kochanek}, \&
  {Stanek}}]{szczygiel12}
{Szczygie{\l}}, D.~M., {Gerke}, J.~R., {Kochanek}, C.~S., \& {Stanek}, K.~Z.
  2012, \apj, 747, 23

\bibitem[{{Tammann} {et~al.}(1994){Tammann}, {Loeffler}, \&
  {Schroeder}}]{tammann94}
{Tammann}, G.~A., {Loeffler}, W., \& {Schroeder}, A. 1994, \apjs, 92, 487

\bibitem[{{Thompson} {et~al.}(2003){Thompson}, {Burrows}, \&
  {Pinto}}]{thompson03}
{Thompson}, T.~A., {Burrows}, A., \& {Pinto}, P.~A. 2003, \apj, 592, 434

\bibitem[{{Thorstensen} {et~al.}(2001){Thorstensen}, {Fesen}, \& {van den
  Bergh}}]{thorstensen01}
{Thorstensen}, J.~R., {Fesen}, R.~A., \& {van den Bergh}, S. 2001, \aj, 122,
  297

\bibitem[{{Timmes} \& {Woosley}(1997)}]{timmes97}
{Timmes}, F.~X., \& {Woosley}, S.~E. 1997, \apj, 489, 160

\bibitem[{{Ugliano} {et~al.}(2012){Ugliano}, {Janka}, {Marek}, \&
  {Arcones}}]{ugliano12}
{Ugliano}, M., {Janka}, H.-T., {Marek}, A., \& {Arcones}, A. 2012, \apj, 757,
  69

\bibitem[{{Vagins}(2011)}]{vagins11}
{Vagins}, M. 2011, in Hamburg Neutrinos from Supernova Explosions, 147

\bibitem[{{Vagins}(2012)}]{vagins12}
{Vagins}, M.~R. 2012, NuPhS, 229, 325

\bibitem[{{van den Bergh}(1975)}]{bergh75}
{van den Bergh}, S. 1975, \apss, 38, 447

\bibitem[{{Vogel} \& {Beacom}(1999)}]{vogel99}
{Vogel}, P., \& {Beacom}, J.~F. 1999, \prd, 60, 053003

\bibitem[{{Walborn} {et~al.}(1987){Walborn}, {Lasker}, {Laidler}, \&
  {Chu}}]{walborn87}
{Walborn}, N.~R., {Lasker}, B.~M., {Laidler}, V.~G., \& {Chu}, Y.-H. 1987,
  \apjl, 321, L41

\bibitem[{{Watanabe} {et~al.}(2009){Watanabe}, {Zhang}, {Abe}, {Hayato},
  {Iida}, {Ikeda}, {Kameda}, {Kobayashi}, {Koshio}, {Miura}, {Moriyama},
  {Nakahata}, {Nakayama}, {Obayashi}, {Ogawa}, {Sekiya}, {Shiozawa}, {Suzuki},
  {Takeda}, {Takenaga}, {Takeuchi}, {Ueno}, {Ueshima}, {Yamada}, {Hazama},
  {Higuchi}, {Ishihara}, {Kajita}, {Kaneyuki}, {Mitsuka}, {Nishino}, {Okumura},
  {Tanimoto}, {Clark}, {Desai}, {Dufour}, {Kearns}, {Likhoded}, {Litos},
  {Raaf}, {Stone}, {Sulak}, {Wang}, {Goldhaber}, {Bays}, {Casper}, {Cravens},
  {Dunmore}, {Griskevich}, {Kropp}, {Liu}, {Mine}, {Regis}, {Smy}, {Sobel},
  {Ganezer}, {Hill}, {Keig}, {Jang}, {Jeong}, {Kim}, {Lim}, {Fechner},
  {Scholberg}, {Walter}, {Wendel}, {Tasaka}, {Guillian}, {Learned}, {Matsuno},
  {Messier}, {Watanabe}, {Hasegawa}, {Ishida}, {Ishii}, {Kobayashi},
  {Nakadaira}, {Nakamura}, {Nishikawa}, {Oyama}, {Sakashita}, {Sekiguchi},
  {Tsukamoto}, {Suzuki}, {Ichikawa}, {Minamino}, {Nakaya}, {Yokoyama},
  {Haines}, {Dazeley}, {Svoboda}, {Gran}, {Habig}, {Fukuda}, {Itow}, {Tanaka},
  {Jung}, {McGrew}, {Sarrat}, {Terri}, {Yanagisawa}, {Tamura}, {Idehara},
  {Ishino}, {Kibayashi}, {Sakuda}, {Kuno}, {Yoshida}, {Kim}, {Yang},
  {Ishizuka}, {Okazawa}, {Choi}, {Seo}, {Furuse}, {Nishijima}, {Yokosawa},
  {Koshiba}, {Totsuka}, {Vagins}, {Chen}, {Deng}, {Gong}, {Liu}, {Xue},
  {Kielczewska}, {Berns}, {Shiraishi}, {Thrane}, {Wilkes}, \& {Super-Kamiokande
  Collaboration}}]{watanabe09}
{Watanabe}, H., {et~al.} 2009, Astroparticle Physics, 31, 320

\bibitem[{{Yakunin} {et~al.}(2010){Yakunin}, {Marronetti}, {Mezzacappa},
  {Bruenn}, {Lee}, {Chertkow}, {Hix}, {Blondin}, {Lentz}, {Bronson Messer}, \&
  {Yoshida}}]{yakunin10}
{Yakunin}, K.~N., {et~al.} 2010, CQGra, 27, 194005

\end{thebibliography}

\end{document}